\documentclass[useAMS,usenatbib]{mn2e}

\usepackage{graphicx}
\usepackage{aas_macros}
\usepackage{placeins}

\usepackage{graphicx}    

\title[Connecting the dots III]{Connecting the dots III: Night side cooling and surface friction affect climates of tidally locked terrestrial planets}
\author[L. Carone, R. Keppens and L. Decin]{L. Carone$^{1}$\thanks{E-mail:
ludmila.carone@wis.kuleuven.be (LC)},  R.
Keppens$^{1}$ and L. Decin$^{2}$\\
$^{1}$Centre for mathematical Plasma Astrophysics, Department of Mathematics, KU Leuven, Celestijnenlaan 200B, 3001 Leuven, Belgium\\
$^{2}$Instituut voor Sterrenkunde, KU Leuven, Celestijnenlaan 200D, 3001 Leuven, Belgium}
\begin{document}

\date{TBD}

\pagerange{\pageref{firstpage}--\pageref{lastpage}} \pubyear{TBD}

\maketitle

\label{firstpage}

\begin{abstract}
We investigate how night side cooling and surface friction impact surface temperatures and large scale circulation for tidally locked Earth-like planets. For each scenario, we vary the orbital period between $P_{rot}=1-100$~days and capture changes in climate states.

We find drastic changes in climate states for different surface friction scenarios. For very efficient surface friction ($t_{s,fric}=$~0.1 days), the simulations for short rotation periods ($P_{rot} \leq$ 10 days) show predominantly standing extra tropical Rossby waves. These waves lead to climate states with two high latitude westerly jets and unperturbed meridional direct circulation. In most other scenarios, simulations with short rotation periods exhibit instead dominance by standing tropical Rossby waves. Such climate states have a single equatorial westerly jet, which disrupts direct circulation.

Experiments with weak surface friction ($t_{s,fric}=~10 -100$ days) show decoupling between surface temperatures and circulation, which leads to strong cooling of the night side. The experiment with $t_{s,fric}=~100$ days assumes climate states with easterly flow (retrograde rotation) for medium and slow planetary rotations $P_{rot}= 12 - 100$~days.

We show that an increase of night side cooling efficiency by one order of magnitude compared to the nominal model leads to a cooling of the night side surface temperatures by 80-100~K. The day side surface temperatures only drop by 25~K at the same time. The increase in thermal forcing suppresses the formation of extra tropical Rossby waves on small planets ($R_P=1 R_{Earth}$) in the short rotation period regime ($P_{rot} \leq$ 10 days).

\end{abstract}



\section{Introduction}
\label{Introduction}
The climate states of tidally locked terrestrial planets do not fail to surprise. \citet{Joshi1997} found that a sufficiently dense atmosphere (i.e., with surface pressure level $p_s \geq 100$~mbar) can maintain generally habitable surface temperatures - even with one planet side permanently facing their host star. Recently, \cite{Yang2014} found that cloud coverage over the substellar point leads to a net cooling of the atmosphere - in contrast to fast rotating non-tidally locked planets. This stabilizing cloud feedback pushes the inner edge of the habitable zone closer towards the star. Even more recently, \cite{Carone2015} (hereinafter called C15) found that fast rotating tidally locked planets ($P_{orb}=P_{rot} \leq 12$~days for large terrestrial planets) may assume up to three different climate states - depending on the dominance of either the tropical or extra tropical standing Rossby wave or a mixture of both.

\cite{Merlis2010} report for $P_{rot}=1$~days two high latitude westerly jets, which we linked in C15 to the dominance of a standing extra tropical Rossby wave. \cite{Edson2011} show for their 'dry model' and the same rotation period a 'mixed' climate state with two high latitude westerly jets and tropical jets, apparently induced by the simultaneous presence of a standing tropical and extra tropical Rossby wave. C15 report for $P_{rot}=1$~days one single equatorial wind jet with very fast wind speeds (about 300 m/s), which we linked to the dominance of a standing tropical Rossby wave.

These three models differ from each other in that \cite{Merlis2010} study the climate of an aqua-planet with moderate day to night side temperature gradient. The authors report maximum surface temperatures of $T_{DS,s}=300$~K at the day side and minimum temperatures of $T_{NS,s}=250$~K at the night side, respectively. \cite{Edson2011} assume for one of their climate models a completely dry surface, which leads to a stronger horizontal surface temperature gradient: They have maximum temperatures of $T_{DS,s}=350$~K at the day side and minimum temperatures of $T_{NS,s}=150-170$~K at the night side, respectively. Our nominal model 'falls in between' in that it has a relatively warm day side with maximum day side surface temperatures of up to $T_{DS,s}=370$~K for $P_{rot}=1$~days and relatively high minimum night side temperatures of $T_{NS,s}=270$~K for an Earth-size planet. Our model, however, has the advantage that we can perform large parametric surveys at low computational cost. Due to the model's transparent and versatile parametrization, we can identify possible sources of deviation and have full control over optical depth, effective radiative time scales on the day and night side, surface friction time scales and the extent of the surface boundary layer.

Indeed, we already identified in \cite{Carone2014} (hereinafter called C14) low night side cooling efficiency as the likely origin for the comparatively warm night side and thus as one source of deviations between our nominal model and that of \cite{Edson2011} and \cite{Joshi2003}.

Another notorious source for deviations between climate models on terrestrial planets are surface boundary prescriptions that vary even for terrestrial Solar System planets by orders of magnitude (C14). We will show, in this study, that different night side cooling efficiencies and, in particular, surface boundary treatments affect climate patterns in the rotation period regime $P_{rot}=1-100$~days.

Due to the computational efficiency of our model, we can perform many simulations for tidally locked planets between $P_{rot}=1-100$~days. We can thus monitor closely changes in climate state transitions with planet rotation. This resolution in rotation period is so far unrivalled. By checking the tropical ($\lambda_R$) and extra tropical Rossby radius of deformation ($L_R$) and by performing a perturbation analysis on the horizontal velocity $\vec{v}$ and geopotential height $z$ (see C15), we can furthermore coherently link climate state phases to tropical and extra tropical Rossby waves.

In the following (Section~2), we will first briefly describe the model, the numerical adjustments for very short orbital periods and introduce the relevant parameter space for tidally locked habitable planets. We will then (Section~3) systematically investigate how an increase of the night side cooling efficiency changes surface temperatures. We will also account for differences in optical depths and planet sizes. We will, furthermore, investigate how more efficient night side cooling affects climate state transitions for specific Rossby radii of deformation over planet size, that is, for $\lambda_R/R_P=1$, $\lambda_R/R_P=0.5$, $L_R/R_P=1$ and $L_R/R_P=0.5$. We will compare our results also to other tidally locked Earth climate models like \citet{Joshi1997,Edson2011,Joshi2003}.

For the investigation of different frictional surface boundary treatments on climate states (Section~4), we use our nominal model on a Super-Earth planet\footnote{We choose $R_P=1.45 R_{Earth}$ instead of $R_P=1.5 R_{Earth}$ to be comparable to C14.} ($R_P=1.45 R_{Earth}$). We study surface friction time scales between $t_{fric}=0.1-100$~days and change the upper extent of the planet boundary layer to values between 70-90\% of surface pressure level $p_s$. These ranges of values were identified in C14 from climate models of Solar System terrestrial planets. For every scenario, we monitor again climate state transitions in Rossby waves for $\lambda_R/R_P=1$, $\lambda_R/R_P=0.5$, $L_R/R_P=1$ and $L_R/R_P=0.5$. We provide in Section~5 a summary of our results and in Section~6 a conclusion and outlook.

This is the first study to coherently investigate different assumptions in night side cooling efficiency and surface friction time scales and how they affect surface temperatures on tidally locked habitable planets over the whole relevant rotation period range. This study will thus provide a better understanding of surface temperatures arising from different climate models for tidally locked terrestrial planets.

\section{Our model}

We introduced in C14 our nominal 3D climate model with simplified forcing that is suitable for a tidally locked terrestrial planet. The model uses Newtonian cooling for the thermal forcing of the dynamical core of the Massachusetts Institute of Technology global circulation model (MITgcm)\footnote{http://mitgcm.org} \citep{Adcroft2004}. MITgcm uses the finite-volume method to solve the primitive hydro-statical equations that can be written as the horizontal momentum, vertical stratification, continuity of mass, equation of state for an ideal gas and thermal forcing equation (See Equations~1-6 in C15).

A first order estimate is used for the radiative time scale $t_{rad}$ in the Newtonian thermal forcing that agrees within one order of magnitude to values derived with full radiative transfer for terrestrial Solar System planets (C14). The details of thermal forcing are further discussed in Section~3. The details of surface boundary treatments, where we use a simple Rayleigh friction scheme for the near-surface part of the atmosphere, are further discussed in Section~4.

\subsection{Numerical setup}
We use 20 vertical levels with equally spaced 50 mbar levels with surface pressure $p_s=1000$~mbar. The C32 cubed-sphere grid is used \citep{Marshall2004}: The sphere is subdivided into six tiles, each with $32\times 32$ elements, thus there are $32\times 32 \times 20$ volume elements per tile. This grid corresponds to a global resolution in longitude-latitude of $128\times 64$ or approximately $2.8^{\circ}\times 2.8^{\circ}$.

As outlined in C14, we initialize the atmosphere with constant temperature T = 264~K and
run for $t_{init} =$~400~days. Data generated before $t_{init}$ are discarded; the simulation is run subsequently for $t_{run}$ = 1000~days and averaged over this time period. The nominal time step is $\Delta t =$~450~s, the smallest investigated planet size, $R_P = 1 R_{Earth}$ , however, requires a reduction to $\Delta t =$~300~s (See also C15).
The horizontal momentum forcing $\mathcal{\vec{F}}_v = \mathcal{\vec{F}}_{fric}+ \mathcal{\vec{F}}_{sponge}$ consists of a surface friction (to be discussed in Section 4) and a sponge layer term that we prescribe as:
\begin{equation}
\mathcal{\vec{F}}_{sponge}=-k_R \vec{v},
\end{equation}
where $k_R$ determines the efficiency of horizontal wind braking in the sponge layer.

The sponge layer stabilizes the upper atmosphere boundary layer against non-physical wave reflection. It may
also be justified physically as a first order representation of gravity wave breaking at low atmosphere pressure. The
sponge layer is located at pressures less than $p=100$~mbar, where $k_R$ increases from zero to $k_{max}$ with decreasing pressure using the prescription of \cite{Polvani2002} (C14). In the nominal set-up, $k_{max}$ was set to 1/80 days${}^{-1}$. Simulations with fast rotations ($P_{rot} \leq 4$~days) show very fast equatorial superrotation at the top of the atmosphere with wind speeds of up to 300 m/s. The very fast winds require a larger efficiency of the upper atmosphere sponge layer. See Table~\ref{tab: tau_sponge} for relevant parameters.

We adopted the dry convection scheme of \citet{Molteni2002} for our model that diffuses dry static energy $s = c_p T + \Phi$ vertically with vertical diffusion time scale $t_{vds}$,  if static stability is locally violated. Here, $c_P$ is specific heat at constant
pressure and $\Phi=gz$ is the geopotential, where $g$ is surface gravity and $z$ is vertical height. In the nominal model and most simulations discussed in this study, the vertical diffusion time scale $t_{vds}=$~1 day is assumed as in C14 and C15.

\begin{table}
\caption{Sponge layer friction time scales used this study}
\begin{tabular}{c|c|}
\hline
$k_{max}$ & $P_{rot}$ \\
{[}days${}^{-1}$] & [days]\\
\hline
1/80 & 5-100 \\
1/40 & 3-4 \\
\hline
\shortstack{1/20 for optical depth $\tau_s=0.62$\\ 1/10 for  optical depth $\tau_s=0.94$} & \shortstack{$\leq 2$ \\ {}}\\
\hline
\end{tabular}
\label{tab: tau_sponge}
\end{table}

\subsection{Investigated parameter space: Tidally locked habitable terrestrial planets around M dwarfs}
\label{sec: planet parameters}

\begin{table}
\caption{Fixed planetary and atmospheric parameters used in this study.}
\begin{tabular}{|l|c|}
Parameter & \\
\hline
Incident net stellar flux $I_{net}$ & 958 W/m${}^2$\\
Obliquity & $0^\circ$\\
$c_p$ & 1.04 J/gK\\
$p_s$ & 1000 mbar \\
Main constituent & $N_2$\\
Molecular mass $\mu$ & 28 g/mol\\
Adiabatic index $\gamma$ & 7/5 \\
Rotation period $P_{rot}$ & 1-100~days\\
\hline
\end{tabular}
\label{tab: Planetary parameters}
\end{table}

We focus on atmosphere dynamics of tidally locked habitable planets around M dwarfs and assume the same set-up as in C14: A nitrogen dominated greenhouse atmosphere with surface pressure $p_s = 1000$ mbar. We further assume that each planet receives the same amount of stellar irradiation as the Earth. See Table~\ref{tab: Planetary parameters} for the relevant planetary parameters.

\cite{Correia2008} and, more recently, \cite{Leconte2015} questioned the assumption that terrestrial planets in the habitable zone of an M dwarf are tidally locked. The climate of a terrestrial planet in the habitable zone must be interpreted, thus, in the context of two extremes in spin-orbit alignment: fast rotating asynchronous rotations like the Earth and tidal-locking. Whereas climate dynamics on fast rotating planets is relatively well understood from the Solar System, tidally locked planets still merit more investigations, as we will show in this study. Furthermore, slow rotating planets like Venus show large similarities in climate patterns to tidally locked states \citep{Yang2014}. In addition, planets at the inner edge of the habitable zone are still probably tidally locked - even in the light of the revised tidal theory of \cite{Leconte2015}. Incidentally, these planets should be the first to become accessible to atmosphere characterization.

We investigate planets of terrestrial composition, where we assume a bulk density of $\rho_{Earth}=5.5$~g/cm${}^3$ for different planet sizes between $R_P=1$ and $1.45 R_{Earth}$. As shown in C15, assuming uniform bulk density makes for easier comparison between radiative and dynamical time scales and thus easier analysis of atmosphere dynamics evolution with planet size. This assumption yields 1 and 3.1 Earth masses for $R_P$ = 1 and $1.45 R_{Earth}$, respectively.

The rotation period range was selected to cover $P_{rot}$ = 1 −- 100 days so that we can compare our results with that
of C15. The investigated rotation period range is at least down to $P_{rot} = 6.5$~days
relevant for the study of the habitable zone around M dwarf stars (\cite{Zsom2013} and \cite{Seager2013}). \cite{Kaltenegger2009} derive $P_{rot}=2-65.5$~days by taking into account M dwarf host stars down to M9V stars, where the authors assumed a constant planet albedo $\alpha=0.3$. C15 derived a relevant orbital period range $P_{orb}=P_{rot}=2-90$~days for stellar masses $M_*=0.1-0.6 M_{sun}$, for which the Earth equivalent incident flux $I_{net}=I_0(1-\alpha)= 958$~W/\boldmath m${}^{2}$ reaches the planet. $I_0=1368$~W/m${}^{2}$\unboldmath is the incident stellar flux for a given orbital period and $\alpha=0,0.3,0.5,0.7,0.9$ was assumed. \cite{Leconte2013} consider $\alpha=0.1-0.5$, which corresponds to a relevant orbital period range between $P_{orb}=6-80$~days. Also, \cite{Yang2013} derive a planetary albedo as high as $\alpha=0.5$ for tidally locked planets around M dwarfs. Table~\ref{tab: Planetary parameters} lists the planetary and atmosphere parameters that we use in this study.

Fast rotation periods ($P_{rot} < $~10 days) allow in any case to study the emergence of dynamical features that are already known from Hot Jupiters. The climate state results for these rotation periods are thus a preparation for a more detailed study of climate dynamics on hot Super-Earth planets that lie in the transition region between terrestrial planets and gas giants
\citep{Madhusudhan2015}.

\subsection{Planetary time and length scales}
\label{sec: time scales}
The introduction of the following time and length scales will be useful in discussing how an increase in night side cooling efficiency and a change in surface boundary prescription affects the overall climate dynamics evolution for different rotation periods.

The dynamical time scale $t_{dyn}$ is defined as \citep{Showmanbook2011}
\begin{equation}
t_{dyn} \propto \frac{R_P}{U}\label{eq:tau_dyn},
\end{equation}
 where U is the mean horizontal velocity.

The radiative time scale used in C14 and C15 is
 \begin{equation}
t_{rad}= \frac{c_p p_s}{4 g \sigma T_s^3}\label{eq:tau_rad},
\end{equation}
where $T_s$ is the local surface temperature, $p_s$ is the surface pressure and $\sigma$ is the Stefan-Boltzmann constant. This first order estimate agrees within one order of magnitude to values derived with full radiative transfer for terrestrial Solar System planets (C14).

C15 confirmed that standing planetary Rossby waves are of uttermost importance for determining climate states of tidally locked planets with rotation periods $P_{rot}$= 1 -- 100 days. More precisely, two types of standing Rossby waves determine which climate state is dominant for specific rotation periods: extra tropical and tropical Rossby waves. The former leads to a climate state with high latitude eastward winds, the latter leads to the formation of equatorial superrotation.

The possible presence of one or both of these Rossby waves can already be inferred from the calculation of the Rossby radii, where $L_R$ denotes the extra tropical Rossby radius and $\lambda_R$ the tropical Rossby radius (see Equations~16 and 17 in C15 or \cite{Holton}). We will identify atmosphere dynamic transitions with respect to $L_R/R_P$ and $\lambda_R /R_P$ in Sections 3.4 and 4.2.

For the identification of the precise type of standing Rossby wave, we will use again the perturbation method as described in Section~4 of C15. Their Figure~2 depicts the expected perturbations in horizontal wind $\vec{v}$ and geopotential height $z=\Phi/g$ at the top of the atmosphere (p = 225 mbar) for the extratropical and extratropical Rossby wave.

\section{Night side cooling adjustment}

We found in C14 that the surface temperatures at the day side of our nominal model were in good
agreement with the Earth climate model of \cite{Edson2011} and \cite{Joshi2003}. Our model's night side temperatures, however, were about 100 K warmer. We speculated in C14 that the night side surface temperatures are mainly determined in our model by the night side cooling efficiency. In the following, we test this hypothesis. We vary the night side cooling efficiency via the night side radiative time scale $t_{rad,NS}$. In addition, we also investigate how changes in optical depth and planet size affect night side surface temperatures. The results of our experiments are also compared to \cite{Joshi1997}, \cite{Joshi2003} and \cite{Edson2011}.

 We use the maximum and minimum surface temperatures to track changes in the day side and night side heating. Surface temperature maxima are always located very close to the substellar point ($\phi=0^{\circ}$) in our climate models. Surface temperature minima are always located at the night side. The latter are, however, shifted from the antistellar point at longitude $\phi=-180^{\circ}$ towards the morning limb, where the displacement increases with faster planet rotation. In the extremest case, surface temperature minima are located at $\phi=-120^{\circ}$. See also Figure~4 of C14 for surface temperature distributions.

\subsection{Radiative efficiency at the night side}

As outlined in C14, we use Newtonian cooling for the temperature forcing  in our model:
\begin{equation}
\mathcal{F}_T=\frac{T-T_{eq}}{t_{rad}},
\end{equation}
where $t_{rad}$ is the radiative time scale as described in Section~2. The equilibrium temperature $T_{eq}$ is described in detail in C14, where we have developed separate prescriptions for the day side and the night side: The illuminated day side is driven towards radiative-convective temperature profiles suitable for a greenhouse atmosphere. The night side temperature relaxes towards the Clausius-Clayperon relation of the main constituent of the atmosphere (nitrogen).

\begin{table}
\caption{Parameters of the experiments used to adjust night side cooling}
\begin{tabular}{l|c|c|c|}
\hline
parameter & Nom.~1 & Exp. 1 & Exp. 2\\
\hline
$R_P$ [$R_{Earth}$] & 1.45& 1.45 & 1.45\\
$g$ [m/s${}^2$] & 14.3 & 14.3 & 14.3\\
$\tau_s$ & 0.62 & 0.62 & 0.94\\
$\alpha$ & 0.3 & 0.3 & 0.3 \\
$T_{s,max}$ [K] & 408 & 408 & 425.5\\
$t_{rad, DS}$ [d]${}^a$ & 5.4${}^b$ & 5.4 & 4.8\\
$t_{rad, NS}$ [d] & 813 & 813-100 & 100\\
\hline
parameter & Nom.~2& Exp. 3  & Exp. 4 \\
\hline
$R_P$ [$R_{Earth}$] & 1 & 1 & 1\\
$g$ [m/s${}^2$] & 9.8 & 9.8 & 9.8\\
$\tau_s$ & 0.62 & 0.94 & 0.94\\
$\alpha$ & 0.3 & 0.3 & 0.3\\
$T_{s,max}$ [K] & 408 & 425.5 & 425.5\\
$t_{rad, DS}$ [d] & 8 & 4.8 &3.3\\
$t_{rad, NS}$ [d] & 1186${}^c$ & 100&69\\
\hline
\end{tabular}
\label{tab: ns_models}
\newline
$a$) $t_{rad, DS}$ is the maximum value at the substellar point.\newline
$b$) Note that in C14 this value was given erroneously as 13~days.\newline
$c$) As discussed in C15
\end{table}

In the first set of simulations (\emph{Exp.1}), we take the nominal model as introduced in C14 as a starting point (\emph{Nom.1}). Keeping all other parameters equal, the night side radiative time scale is successively lowered from $t_{rad,NS}$ from 813 days to $t_{rad,NS}$=100~days (Table~\ref{tab: ns_models}).

\begin{figure}
\includegraphics[width=0.48\textwidth]{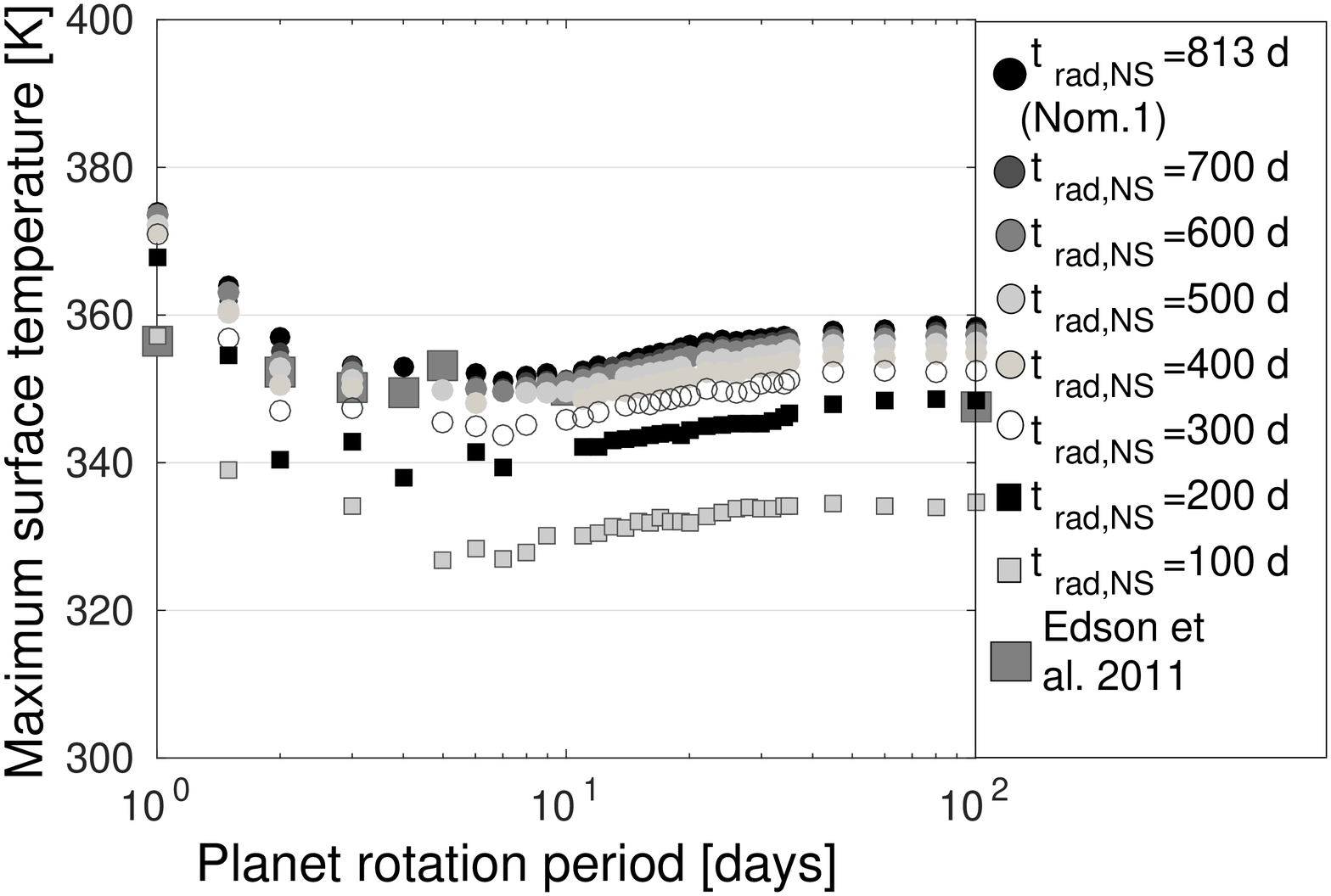}
\includegraphics[width=0.48\textwidth]{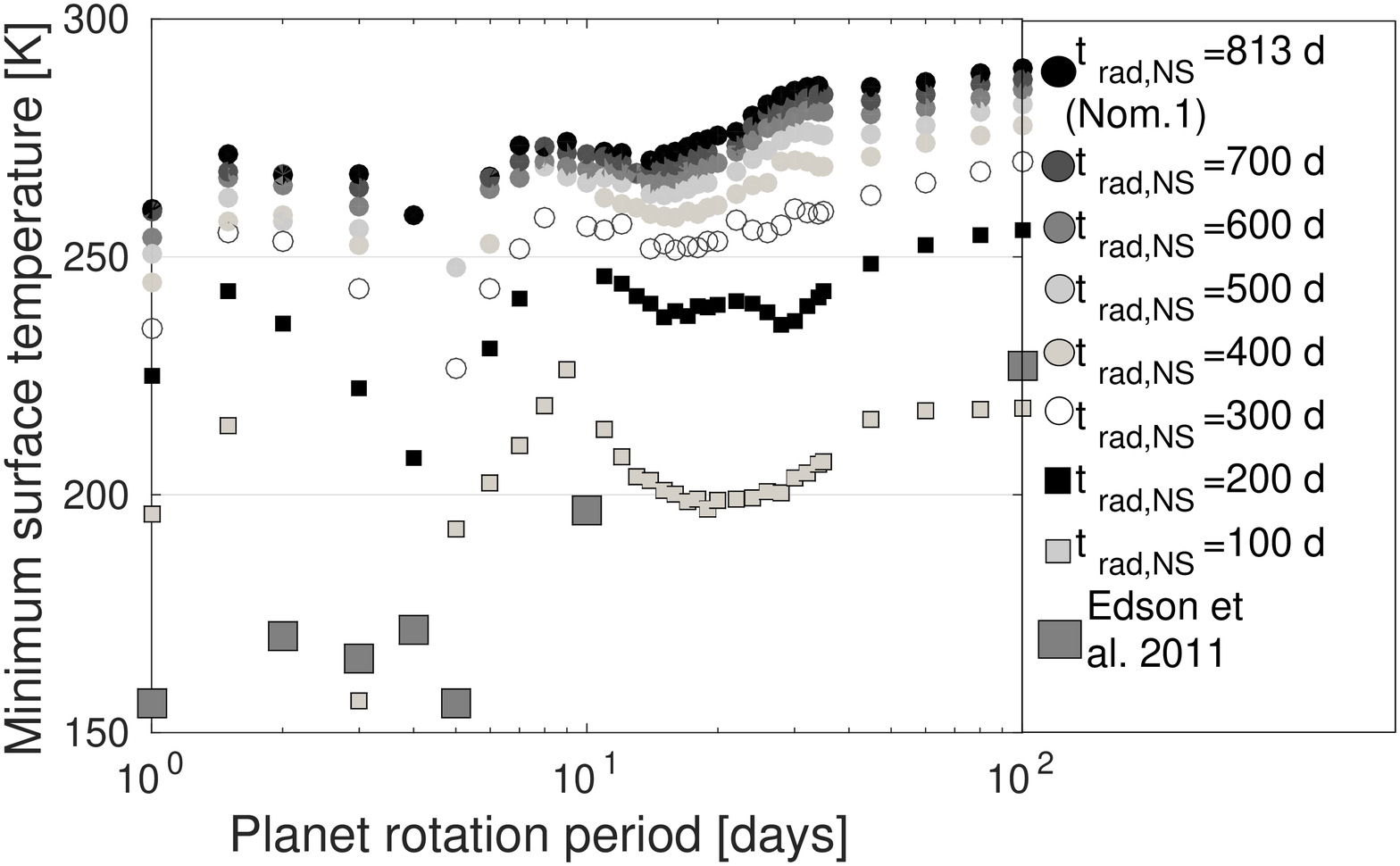}
\caption{Comparison between maximum (top) and minimum surface temperatures (bottom) in \textit{Nom.1} and \textit{Exp.1} for different nightside cooling time scales $t_{rad,NS}$. For comparison, the surface temperatures reported by \citet{Edson2011} are shown.}
\label{fig: surf_nom}
\end{figure}

Figure~\ref{fig: surf_nom} shows that just reducing the night side radiative time scale to $t_{rad,NS}$ = 100 days, reduces the surface temperatures at the night side by about 80-100 K. The day side surface temperatures also cool down but to a much lesser degree - by just 25~K compared to the nominal model (\emph{Nom.1}). \emph{Exp.1} with $t_{rad,NS}$ = 100 days shows furthermore good agreement to night side temperatures from the dry model of \cite{Edson2011} in the slow rotation regime $P_{rot}\geq 4$~days. The day side temperatures are colder by 10-20~K.

In the fast rotation regime with $P_{rot}\leq 4$ days, the comparison in surface temperatures between our model and other models is confounded by different possible climate states as demonstrated in C15: Whereas the model of \cite{Edson2011} assumes a climate state with two high latitude westerly wind jets due to the dominance of a standing extra tropical Rossby wave, our model adopts a climate state with a single equatorial superrotating jet due to the dominance of the standing tropical Rossby wave. The latter state suppresses at least the meridional part of the equatorial direct circulation cell and thus suppresses efficient cooling of the substellar point via upwelling. The suppression of equatorial circulation in \textit{Exp.1} and \textit{Nom.1} leads to very warm substellar point temperatures for very short rotation periods $P_{rot} \leq 4$~days - in contrast to the model used by \cite{Edson2011} that assume colder day side temperatures due to unperturbed direct circulation.

Despite stronger suppression of direct circulation for faster rotation in the $P_{rot} \leq 4$~days-regime, the night side surface temperature in \textit{Exp.1} increase with faster rotation. As outlined in C15, the rise in night side surface temperatures with faster rotation is attributed to the increasing strength of the secondary circulation cells that transfer heat efficiently from the day side towards the night side surface.

In general, we can confirm that it is indeed mainly the night side cooling efficiency that determines night side surface temperatures in our simplified model prescription. At the same time, a reduction in $t_{rad,NS}$ effects the day side surface temperatures only moderately and apparently does not change climate states compared to \textit{Nom.1}(discussed in C15).

\subsection{Adjustment for differences in optical depth}

\begin{table}
\caption{Vertical diffusion time scales of the experiments used to adjust night side cooling}
\begin{tabular}{c|c|}
\hline
$t_{vds}$ & $P_{rot}$ \\
{[}days] & [days]\\
\hline
1 & 10-100 \\
\hline
\shortstack{1 \\for Nom.1, Nom.2, Exp.1, Exp.3
 \\0.8 \\ for others} & \shortstack{{}\\{9-4}\\{}\\{}}\\
\hline
\shortstack{1 \\for Nom.1, Nom.2, Exp.1 \\0.8\\ for others} & \shortstack{{}\\{3-1.5}\\{}\\{}}\\
\hline
\shortstack{1 \\for Nom.1, Nom.2, Exp.1\\0.5 for Exp.2\\0.2 for Exp.3\\0.1 for Exp.4} & \shortstack{{}\\{1}\\{}\\{}\\{}}\\
\hline
\end{tabular}
\label{tab: tau_vds}
\end{table}

Another complication that arises when comparing different climate models are differences in optical depth, even assuming the same atmosphere composition. More precisely, $\tau_s$=~0.62 was assumed in our nominal model and \textit{Exp.1}. \cite{Edson2011} adopt $\tau_s$=~0.94 (their Table~4). To investigate changes in optical depth, we performed another experiment (\emph{Exp.2}) with $t_{rad,NS}$=~100 days but with day side equilibrium temperatures calculated for $\tau_s$=~0.94 (Table~\ref{tab: ns_models}).

\begin{figure}
\includegraphics[width=0.48\textwidth]{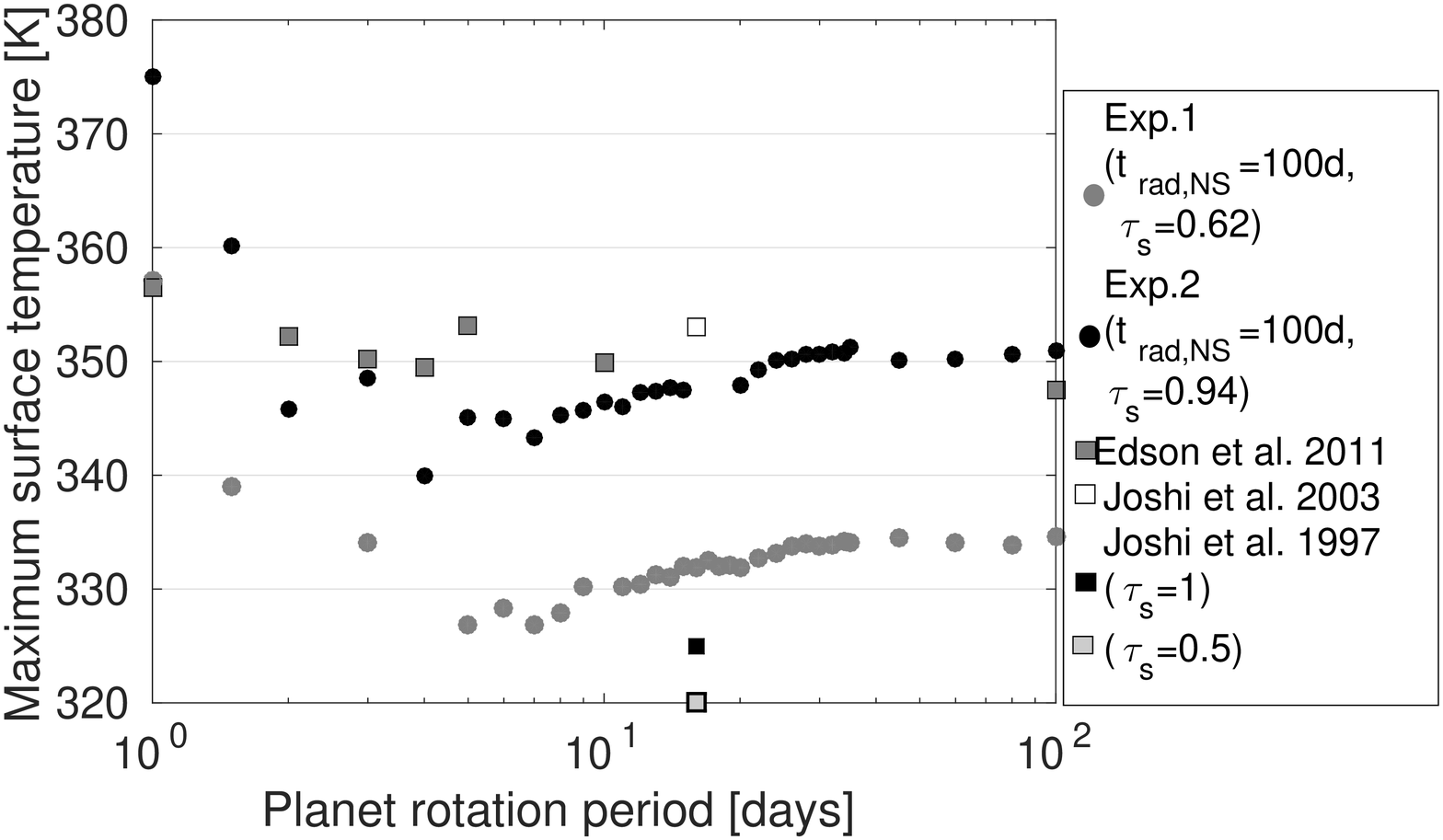}
\includegraphics[width=0.48\textwidth]{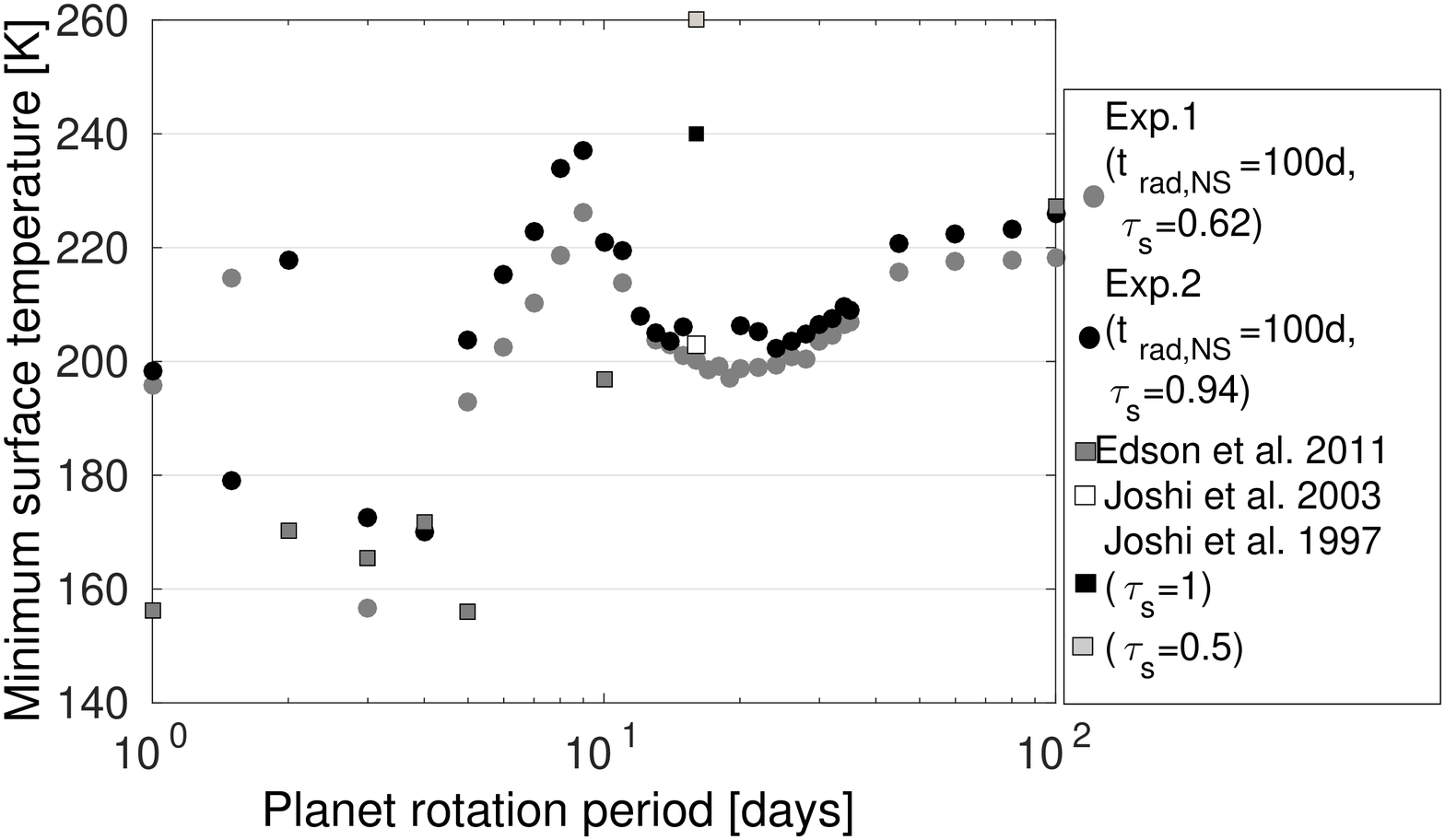}
\caption{Comparison between maximum (top) and minimum (bottom) surface temperatures in Experiments with different optical depths (\textit{Exp.1} and \textit{2}). For comparison, surface temperatures reported by \citet{Edson2011}, \citet{Joshi2003} and \citet{Joshi1997} are shown.}
\label{fig: surf_tau}
\end{figure}

Figure~\ref{fig: surf_tau} shows that an increase in optical depth leads to higher day side surface temperatures, while the night side remains virtually unaffected. The surface temperature dependencies with optical depth in our prescription as demonstrated by \textit{Exp.2} are very different to the results reported by \cite{Joshi1997}, who performed similar experiments by changing optical depth between $\tau_s$ = 0.25 -- 2, where only results for 0.5 and 1 are compared in this study. The change in optical depth led there to hardly any changes in day side surface temperatures but affected instead the night side temperatures very strongly.

In addition, we find that the stronger heating of the day side surface in our experiments with higher optical depth (\textit{Exp.2}, \textit{Exp.3} and \textit{Exp.4}) triggers very strong upwelling over the substellar point. The stronger upwelling leads to violation of static stability for short rotation periods ($P_{rot}\leq 9$~days), if the nominal dry adiabatic adjustment time scale $\tau_{vds}=1$~days is used (Section~2.1). When a simulation showed instability, we reduced the diffusion time scale $\tau_{vds}$ in 0.1~s steps, until static stability could be established again. We find that \textit{Exp.4} with strong night side cooling requires generally the smallest $t_{vds}$ in the short rotation regime. We thus justify dry adiabatic adjustment with shorter diffusion time scale as a parametrization of more vigorous upwelling, which is the strongest for \textit{Exp.4}. The necessary adjustments in vertical diffusion time scales $t_{vds}$ are listed in Table~\ref{tab: tau_vds}.

\subsection{Adjustment for planet size}
\label{sec: Adjust_planet_size}

\begin{figure}
\includegraphics[width=0.48\textwidth]{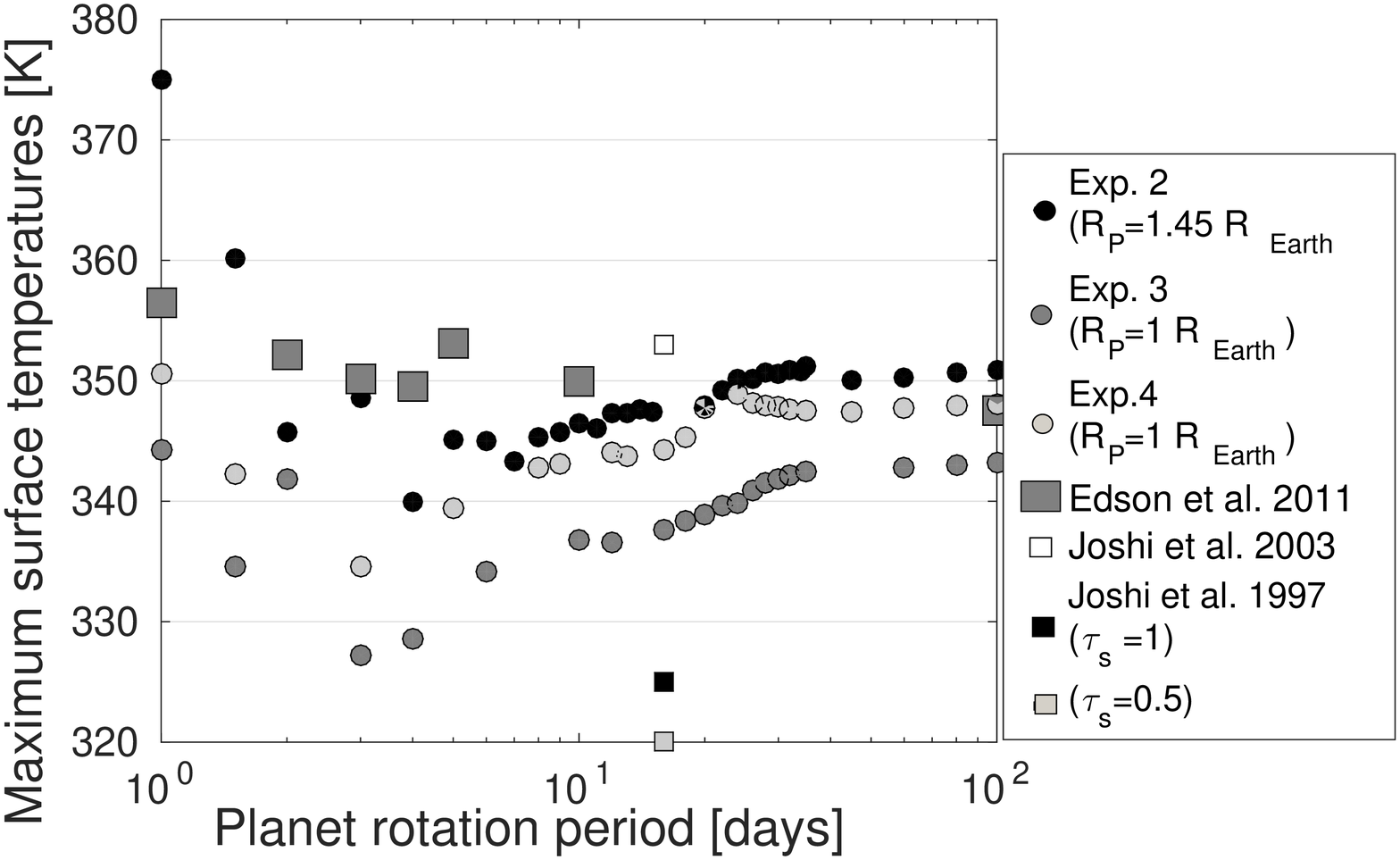}
\includegraphics[width=0.48\textwidth]{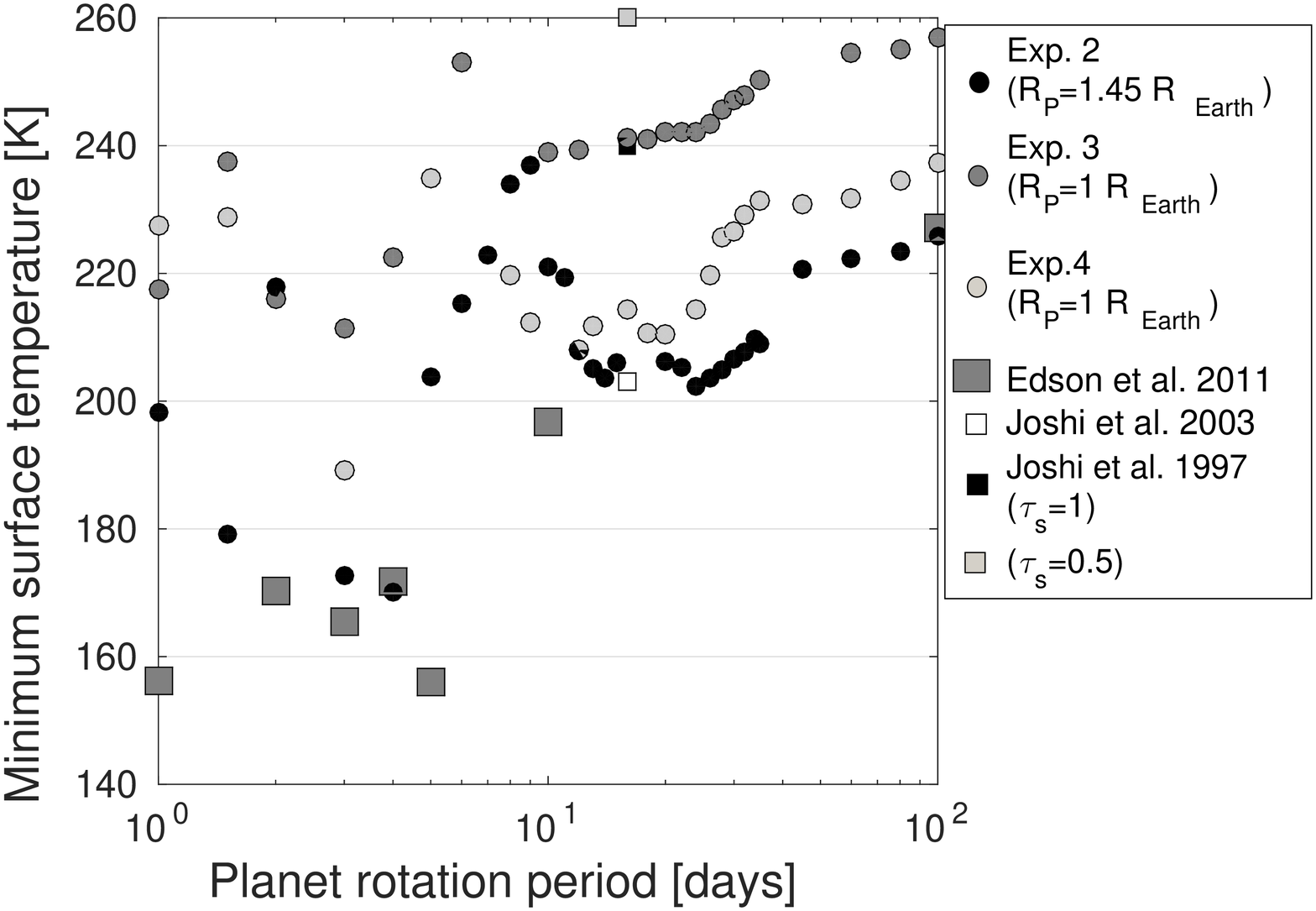}
\caption{Maximum (top) and minimum (bottom) surface temperatures in \textit{Exp.2}, \textit{Exp.3} and \textit{Exp.4} that show adjustment in $t_{rad}$ to compensate for increase in dynamics, when changing the planet size from $R_P=1.45 R_{Earth}$ to $R_P= 1 R_{Earth}$. In addition, surface temperatures reported by \citet{Edson2011}, \citet{Joshi2003} and \citet{Joshi1997} are shown for comparison. }
\label{fig: surf_ave_Rp}
\end{figure}

Up to now, we investigated night side cooling variations for terrestrial-like atmosphere on a tidally locked Super Earth planet with radius $R_P = 1.45 R_{Earth}$  (\textit{Exp.1} and \textit{Exp.2}). Many other tidally locked Earth climate models assume, however, $R_P = 1 R_{Earth}$ like e.g. \cite{Edson2011}, \cite{Joshi1997} and \cite{Joshi2003}. It was shown in C15 that just changing the planet size within our model formulation already leads to changes in the gradient between day side to night side surface temperatures. These changes in planet size were connected to the change in dynamical time scale $t_{dyn}$, where $t_{dyn} \propto R_P$ (Equation~\ref{eq:tau_dyn}). In C15, however, also the radiative time scales $t_{rad}$ changed via surface gravity $g$ (Equation~\ref{eq:tau_rad}): $t_{rad}$ was increased with decreasing planet size. This nominal experiment for an Earth-size planet without increased night side cooling efficiency (\textit{Nom.2}) was investigated in detail in C15. The corresponding parameters are shown in Table~\ref{tab: ns_models}.

For \textit{Exp. 3}, we reduce the planet size compared to \textit{Exp.2}, but keep $\tau_{rad}$ unchanged in contrast to C15. That is, we adopt $R_P = 1 R_{Earth}$ and $g=9.8$~m/s${}^2$ with the same radiative time scales as in \textit{Exp.2}. Any changes that arise between \textit{Exp.2} and \textit{Exp.3} can be therefore exclusively linked to changes in the dynamical time scale $t_{dyn}$ with reduced planet size.

Figure~\ref{fig: surf_ave_Rp} shows that just reducing the planet size - keeping all other parameters equal - leads to a decrease in surface temperature day side to night side gradient compared to \textit{Exp.2}. In particular, the night side temperatures are substantially higher in \textit{Exp.3}. This result is in accordance to C15, where it was likewise found that the gradient between night side to day side becomes smaller for smaller planet sizes. In C15, this effect was attributed to the combined effect of an decrease in dynamical time scale $\tau_{dyn}$ and an increase in radiative time scale. In other words, we concluded that dynamics becomes more efficient on smaller planets, whereas radiative heating becomes weaker.

Like in C15, we now assume that dynamics - in the form of flow from the day side towards the night side and vice versa- is mainly driven by temperature gradient between the day and night side, that in turn is maintained by radiative heating rate $J$, the efficiency of which is prescribed in our simplified forcing via $t_{rad}$:
\begin{equation}
J=\frac{\Delta \theta} {t_{rad}},\label{eq: heating_rate}
\end{equation}
where $\Delta \theta$ is the difference between the prescribed potential equilibrium temperature and the atmosphere potential temperature. Applying scale analysis on the thermodynamic energy equation (like in C15, Section 5.1.1), the radiative heating rate $J$ can be linked to dynamics via:
\begin{equation}
\frac{\theta U} {R_P} \sim \frac{J} {c_P},
\end{equation}
where $\theta(x,y)$ is in the context of tidally locked terrestrial planets the actual difference between day to night side temperatures. Inserting Equation~(\ref{eq:tau_dyn}), we can link $t_{dyn}$ to heating rate and thus $t_{rad}$ via:
\begin{equation}
J \sim \frac{\theta}{t_{dyn}}.
\end{equation}

Because the heating rate $J$ did not substantially change between \textit{Exp.2} and \textit{Exp.3}, we assume that differences in $\theta$ are mainly driven by $t_{dyn}$. We furthermore assume that we can adjust the heating rate $J$ via $t_{rad}$  to compensate for the more efficient dynamics, increasing the temperature gradient between day side and night side again: We  decrease $t_{rad}$ at the day and night side by a factor of $1.45$ (Table~\ref{tab: ns_models}) to compensate for changes in $t_{dyn}$ (\textit{Exp.4}).\textit{ Exp.4} implies that $t_{dyn}\propto R_P$ and that $U$ is to first order constant between \textit{Exp.3} and\textit{Exp.4}. Figure~\ref{fig: surf_ave_Rp} shows that these assumptions are apparently valid: The surface temperatures of \textit{Exp.4} are in better agreement with \textit{Exp.2} than \textit{Exp.3}. This relative agreement is particularly evident in the slow rotation regime $P_{rot}\geq 10$~days.

The climate states did not change between \textit{Exp.2} and \textit{Exp.4}, even in the short planet rotation period regime $P_{rot}=1-3$~days, where we notice the strongest deviations in between our nominal model and other tidally locked Earth models like \cite{Edson2011} (C15).

It was shown in C15 that the latter deviations arise because three climate states are possible for $P_{rot} \leq 3$~days on an Earth-size planet: one dominated by standing tropical Rossby waves leading to fast equatorial superrotation, one dominated by extra tropical Rossby waves and high latitude westerly jets and one 'mixed' climate state with contributions from both the tropical and extra tropical Rossby wave and thus a mixture between high latitude westerly jet with equatorial superrotation. The first climate state with pure equatorial superrotation leads to inefficient cooling of the substellar point in contrast to the other two climate states. Its presence can thus already be inferred from day side surface temperatures that rise steeply with faster rotation. This climate state is apparently adopted by \textit{Exp.2}, \textit{Exp.3} and \textit{Exp.4} for $P_{rot}=1-3$~day (Figure~\ref{fig: surf_ave_Rp}).

\textit{Exp.3} and \textit{Exp.4}, thus, deviate from the climate states of the corresponding nominal model \textit{Nom.2} in the fast rotation regime. It was found in C15 that the smallest investigated planets ($R_P=1-1.25 R_{Earth}$), including \textit{Nom.2}, show climate states with strong contribution of the extra tropical Rossby wave with high latitude wind jets and only weak or no equatorial superrotation. The efficient night side cooling in \textit{Exp.3} and \textit{Exp.4} favours the formation of standing tropical Rossby waves in our model prescription.

Using \textit{Exp.4}, we can also now more coherently compare surface temperatures arising from our model prescription with increased night side cooling efficiency with other tidally locked terrestrial climate models. For $P_{rot} = 10 - 100$~days, \textit{Exp.4} agrees with the results of \cite{Edson2011} and \cite{Joshi2003} within 20 K for both the day side and the night side (Figure~\ref{fig: surf_ave_Rp}). There are, however large deviations between \textit{Exp.4} and \cite{Joshi1997}: The day side surface temperatures are higher by 25-30~K in the former compared to the latter. The night sides are 30-50~K lower in \textit{Exp.4} as compared to \cite{Joshi1997}.

For $P_{rot}$ = 4 -- 10 days, the agreement between surface temperatures at the night side between \textit{Exp.4} and \cite{Edson2011} is less good. \textit{Exp.4} is generally warmer by 30-80~K than the dry model of \cite{Edson2011}. This deviation is, however, small compared to the 100 K difference between \cite{Edson2011} and our nominal model (C14,C15). For even faster rotations ($P_{rot}\leq $ 4 days), deviations in the day side surface temperatures arise because \textit{Exp.4} and the model of \cite{Edson2011} are in different climate states. The short rotation period simulations shown by \cite{Edson2011} are in climate states dominated by standing extra tropical Rossby waves with high latitude westerly jets and efficient direct circulation even for $P_{rot}=1$~day (see also C15). The efficient direct circulation in the simulations of \cite{Edson2011} prevents the substellar point from reaching surface temperatures hotter than 360~K, even when $P_{rot}=1$~day (Figure~\ref{fig: surf_ave_Rp}).

It is striking that despite better agreement in surface temperatures between \textit{Exp.4} and the model of \cite{Edson2011}, compared to the nominal model (\textit{Nom.2}), the disagreement between climate states in the short rotation period regime is even stronger: While \textit{Nom.2} exhibited at least sometimes standing extra tropical Rossby waves for $P_{rot} \leq 3$~days like the model of \cite{Edson2011}, these tendencies disappear in \textit{Exp.4} with stronger night side cooling. Apparently, tropical Rossby waves become stronger when the thermal forcing increases in our model and dominate in \textit{Exp.4} for all rotation periods.

The comparison between experiments with efficient and inefficient night side cooling and with other climate models in the fast planet rotation regime demonstrates an important principle: It is vital to first understand the possible basic climate states arising from 3D climate models, before attempting to compare their details with each other.

\subsection{Effect of increased night side cooling efficiency on large scale climate dynamics}

Up to now, we have only inspected the evolution of surface temperatures between $P_{rot}=1-100$~days. Based on this evolution we already concluded that an increase in night side cooling efficiency not only leads to a strong cooling of the night side, but also favours more strongly the formation of tropical Rossby waves - even for small planets ($R_P=1 R_{Earth}$) and fast planet rotations ($P_{rot}\leq 3$~days).

In the following, we  aim to confirm the conclusion that climate states brought upon by standing tropical Rossby waves are more prevalent in simulations with efficient night side cooling, represented by \textit{Exp.4}, compared to the nominal model (\textit{Nom.2}). For this purpose, we investigate the evolution of climate states for $P_{rot}=1-100$~days by consulting additional diagnostics: the evolution of zonal wind versus rotation period, Rossby wave numbers, and circulation patterns. We will discuss deviations from climate states compared to \textit{Nom.2} that are described in detail in C15.

\begin{figure}
\includegraphics[width=0.48\textwidth]{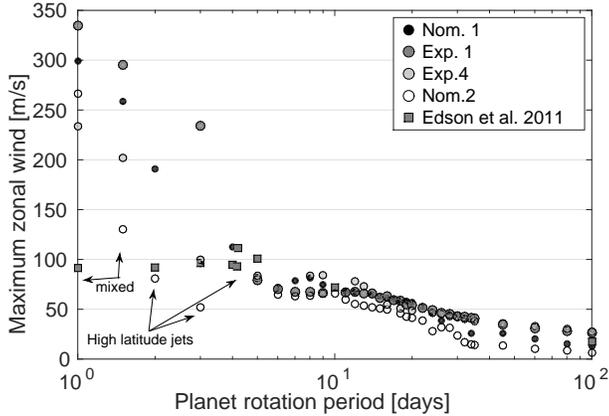}
\caption{Maximum of zonal wind speeds in our model with efficient and inefficient night side cooling, different optical depths and planetary radii and as reported by \citet{Edson2011} in their model.}
\label{fig: zonalwind}
\end{figure}

\subsubsection{Zonal wind}
The inspection of the maximum zonal wind speeds versus planet rotation periods is a good starting point to discuss climate dynamics in a
large parameter study as demonstrated in C15 and \cite{Edson2011}. In this study, we find that wind speeds arising from simulations with increased night side cooling efficiency (\textit{Exp.1} and \textit{Exp.4}) are generally larger than the respective nominal simulations with less efficient night side cooling (\textit{Nom.1} and \textit{Nom.2}) (Figure~\ref{fig: zonalwind}). The larger wind speeds arise because \textit{Exp.1} and \textit{Exp.4} have smaller radiative time scales, increasing the heating rate via $J \propto t_{rad}^{-1}$, which triggers faster winds.

For faster planet rotations ($P_{rot} \leq$~3 days), \textit{Exp.1} and \textit{4} exhibit fast equatorial superrotation, whereas \textit{Nom.2} shows climate states with high latitude wind jets or at least a mixed state for $P_{rot}-1.5,2$ and 3 days, with both equatorial superrotation and high latitude wind speeds existing side by side (C15). High latitude jets have much slower wind speeds than those of superrotating equatorial jets, which explains why \textit{Nom.2} has slower wind speeds in the short planet rotation regime compared to \textit{Exp.4} (Figure~\ref{fig: zonalwind}). The zonal wind evolution thus confirms that more efficient night side cooling in our model prescription and thus stronger thermal forcing suppresses the formation of standing extra tropical Rossby waves and thus the formation of high latitude wind jets also for small exoplanets.

Interestingly, \textit{Exp.4} yields zonal wind speeds in the intermediate and slow rotation period range $P_{rot}=$ 4-100 days comparable to those reported by \cite{Edson2011} (Figure~\ref{fig: zonalwind}). \textit{Exp.4} assumes the same optical depth and planet size as the dry climate model used by \cite{Edson2011}. This agreement indicates that the thermal forcing between \textit{Exp.4} and the model used by \cite{Edson2011} are of similar strength. The similarity in thermal forcing was already inferred in the previous Section by similar gradients in the day side to night side surface temperatures. This result confirms once again that the climate states arising in the fast rotation regime can be very different, despite similar strength in thermal forcing between \textit{Exp.4} and the model of \cite{Edson2011}.
\begin{figure}
\includegraphics[width=0.48\textwidth]{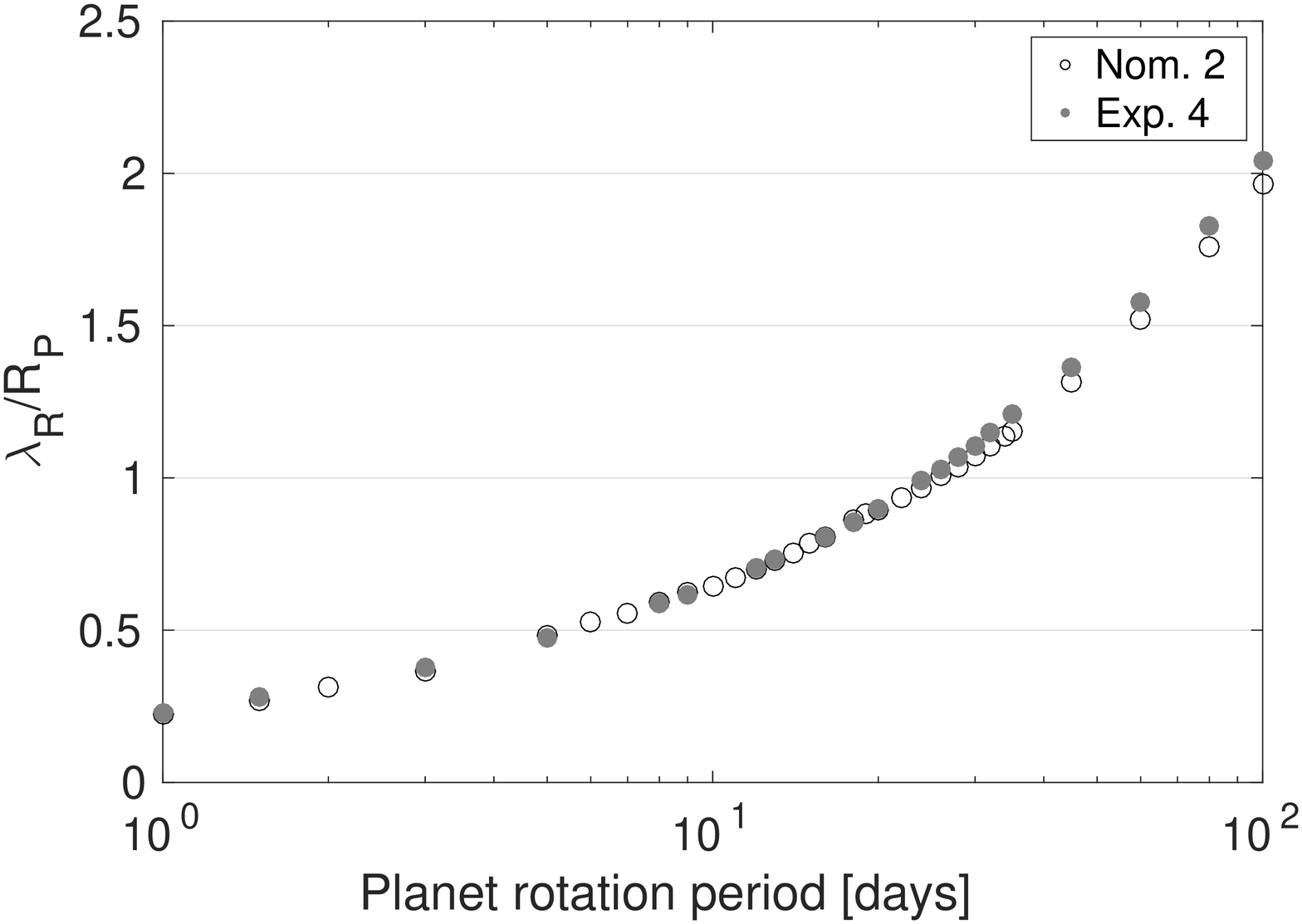}
\includegraphics[width=0.48\textwidth]{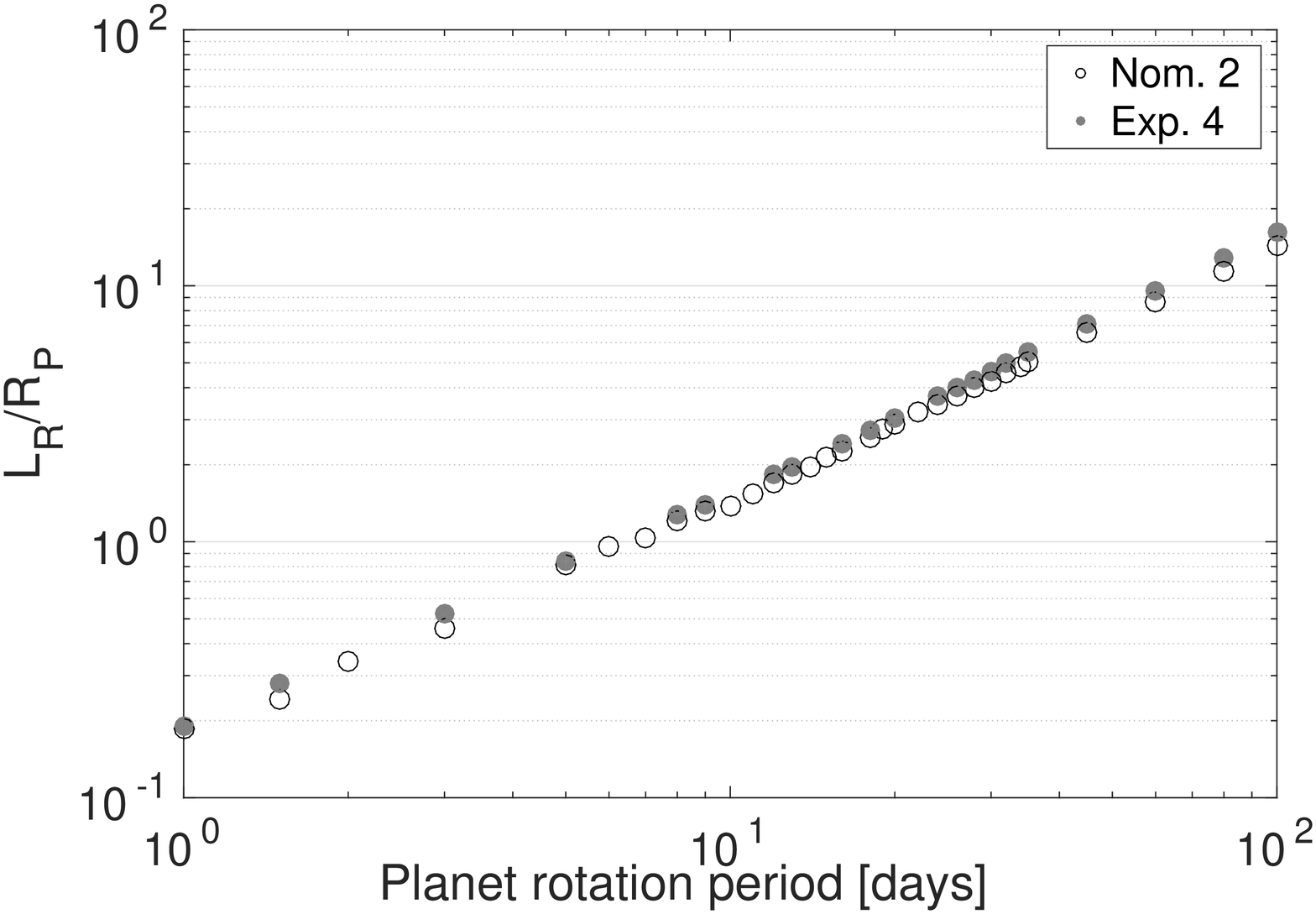}
\caption{Tropical Rossby radius of deformation $\lambda_R$ (top) and extratropical Rossby radius of deformation over planet size $R_P=1 R_{Earth}$ (bottom) for \textit{Exp.4} and \textit{Nom.2}.}
\label{fig: lambda}
\end{figure}

As reported in C15, there are distinct phase changes in climate states for planet rotation periods between 1 and 100 days that can be linked to the tropical and extra tropical Rossby radii of deformation over planet size, $\lambda_R/R_P$ and $L_R/R_P$, respectively. Figure~\ref{fig: lambda} shows that all climate state transitions connected to the tropical and extra tropical Rossby wave are shifted slightly towards faster planet rotation for \textit{Exp.4} compared to \textit{Nom.2}. The concerned climate states are those for which either the tropical or extra tropical Rossby wave becomes smaller than the planetary radius ($\lambda_R /R_P \leq$ 1 and  $L_R /R_P \leq$ 1, respectively) and those for which either the tropical or extra tropical Rossby wave becomes smaller than half the planetary radius ($\lambda_R /R_P \leq$ 0.5 and  $L_R /R_P \leq$ 0.5, respectively). Closer inspection (not shown) shows that the shift is due to an increase in scale height $H$ of \textit{Exp.4} compared to \textit{Nom.2} due to the higher average surface temperatures.

We now monitor the presence of planetary waves for different climate states with respect to the tropical and extra tropical Rossby radii over planet size:
\begin{itemize}
\item $\lambda_R/R_P > 1$,
\item $\lambda_R/R_P \approx 1$,
\item $L_R/R_P \leq 1$ and $\lambda_R/R_P \leq 0.5$  and
\item $L_R/R_P \leq 0.5$ .
\end{itemize}
 For this purpose, we use the perturbation method and inspect the eddy geopotential height $z'$ and the eddy horizontal wind velocities $v'$ on top of the atmosphere ($p=225$~mbar) for planet rotation periods around the relevant Rossby wave numbers.

\subsubsection{Slow rotations with $\lambda_R/R_P > 1$}
\label{sec: ns_slow}

\begin{figure}
\includegraphics[width=0.48\textwidth]{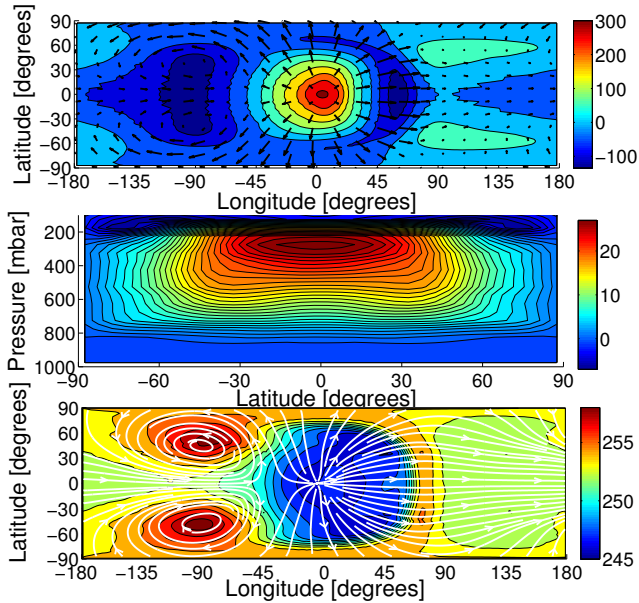}
\caption{The climate state of the $P_{rot}=100$~days simulation for \textit{Exp.4}. Top: Eddy geopotential height in [m] and eddy horizontal wind in [m/s] at p = 225 mbar. Middle: Zonal mean of the zonal flow in [m/s]. Bottom:  Temperatures (contour interval 1 K) and streamlines of the horizontal flow at p=225 mbar. The longest eddy wind vectors are 64.6~m/s. Contour intervals are 50 m for the eddy geopotential and 1 m/s for zonal winds, respectively.}
\label{fig: Prot100d}
\end{figure}

Interestingly, we find that even for the slowest rotator with $P_{rot}$ = 100 days, there are changes in climate dynamics.
\textit{Exp.4} is no longer exclusively dominated by direct divergent flow from the substellar point towards the night side in contrast to \textit{Nom.2} (C15). Divergent flow is still the main feature in this rotation regime as evidenced by the positive geopotential perturbation associated with divergent flow over the substellar point (Figure~\ref{fig: Prot100d}, top). This flow leads, as described in C15, to relative cold upper atmosphere temperatures (Figure~\ref{fig: Prot100d}, bottom). However, in addition two cyclones with local temperature maxima on the north-west and south-west flank of the substellar point (longitude $\phi=-90^{\circ}$) are visible (Figure~\ref{fig: Prot100d} top and bottom). These cyclones are associated with negative geopotential perturbations between latitudes $\pm 50^{\circ}$.

Geopotential perturbations at the equator are the hallmark of a Kelvin wave \citep{Matsuno1966}. The two cyclones and the east-west asymmetry in eddy geopotential height and flow was also observed around the Kelvin wave by \citet{Gill1980}. At the same time, it appears that the negative geopotential height perturbation at the equator ($\phi=-90^{\circ}$) is superimposed by two negative geopotential perturbations with Rossby wave gyres centred at latitudes $\pm 40^{\circ}$ (See also C15, Figure~2 bottom). Shear between Kelvin waves and tropical Rossby waves typically causes equatorial superrotation, which is indeed observed in \textit{Exp.4} (Figure~\ref{fig: Prot100d}, middle panel).

The presence of a stronger Kelvin wave compared to the nominal model \textit{Nom.2} may be explained by the stronger thermal forcing: Kelvin waves are associated with longitudinally varying heating patterns and we increased the temperature difference between day side and night side in \textit{Exp.4} compared to \textit{Nom.2}. On the other hand, the dry model of \cite{Edson2011} shows purely divergent flow for $P_{rot}=100$~days without any indication of Kelvin waves, even though their model has similar strength in thermal forcing than \textit{Exp.4}. Thus, we conclude that strong thermal forcing is one, but not the sole, factor that can trigger Kelvin waves even for slow planet rotation.

The appearance of tropical Rossby gyres in \textit{Exp.4} already for $P_{rot}=100$~days is, however, puzzling.  We have found in C15 the formation of tropical Rossby waves only for $\lambda_R/R_P \leq 1$. Here,  $P_{rot}=100$~days corresponds to $\lambda_R/R_P =2$ (Figure~\ref{fig: lambda}). Just based on the tropical Rossby radius of deformation, we would not expect standing Rossby waves.

The Rossby wave, however, can only be weak: Equatorial superrotation in \textit{Exp.4} for $P_{rot}=100 $~days is relative slow ($\approx 25$~m/s). Therefore, the horizontal flow in the upper atmosphere is still largely dominated by direct divergent flow as evidenced by the velocity streamlines in the upper atmosphere (Figure~\ref{fig: Prot100d}, middle). The corresponding geopotential height anomalies associated with possible Rossby wave gyres are also much smaller than the positive geopotential perturbation over the substellar point associated with the hot substellar point (Figure~\ref{fig: Prot100d}, top). We thus conclude that the additional appearance of weak Kelvin and Rossby waves has in this specific example only a minor influence. Direct divergent flow still dominates in this slow rotation regime, as expected. We can conclude that \textit{Exp.4} shows not only a stronger tendency to form tropical Rossby waves in the short rotation period but also, surprisingly, in the long rotation period regime ($P_{rot}=100$~days) compared to \textit{Nom.2}.

\begin{figure*}
\includegraphics[width=0.9\textwidth]{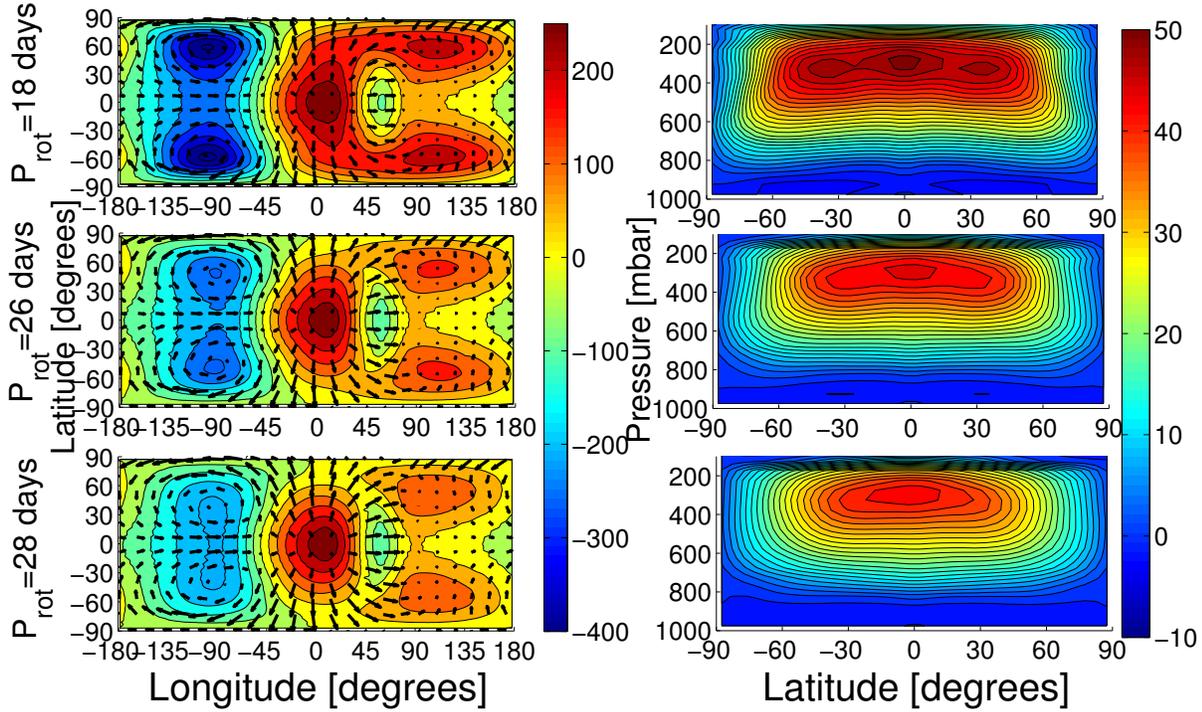}
\caption{The climate state transition, $\lambda_R/R_P \approx 1$, and \textit{Exp.4} for $P_{rot}=18,26$ and 28~days (from top to bottom). Left: Eddy geopotential height in [m] and eddy horizontal wind in [m/s] at p = 225 mbar. Right: Zonal mean of the zonal flow in [m/s]. The longest eddy wind vectors are from top to bottom 64.3, 64.0 and 56.1~m/s, respectively. Contour intervals are 50 m for the eddy geopotential and 2 m/s for zonal wind speeds, respectively.}
\label{fig: lambda1}
\end{figure*}

\subsubsection{Intermediate rotations with $\lambda_R/R_P \approx 1$}
\label{sec: ns_lambda1}

The climate state transition at $\lambda_R/R_P \approx 1$ towards atmosphere dynamics with a standing tropical Rossby wave was identified in C15 with the emergence of geopotential height anomalies away from the equator (see C15, Figure~1, bottom). While this transition is straightforward and easy to identify for models that are exclusively dominated by direct circulation flow in the slow rotation regime (\textit{Nom.2}), it is interesting to observe what happens in \textit{Exp.4}.

While weak Rossby waves are already present for relative slow rotations in \textit{Exp.4} (see previous Section), the waves only fully form when the tropical Rossy wave number becomes smaller than unity. The transition to a climate state that is dominated by standing tropical Rossby waves can be identified with two distinct negative geopotentials perturbations at longitude $\phi=-90$ and latitudes $\nu=\pm 60$. These start to form two distinct gyres for $P_{rot}=26$~days and faster rotations. This transition occurs indeed at $\lambda_R/R_P = 1$ (Figure~\ref{fig: lambda}). Furthermore, the formation of Rossby wave gyres leads to an increase in equatorial wind speeds with faster and faster rotation  (Figure~\ref{fig: lambda1}, right panel at $p=300$~mbar). The full formation of standing tropical Rossby waves in the expected rotation regime confirms that the planetary waves reported in the previous section can only be weak.

\subsubsection{Climate state bifurcation I: $L_R/R_P \leq 1$ and $\lambda_R/R_P \leq 0.5$ }

As reported in C15, the transition in extra tropical Rossby wave via $L_R/R_P \leq 1$ and in tropical Rossby via $\lambda_R/R_P \leq 0.5$ can occur roughly at the same rotation period. This is, $P_{rot}=6$~days for \textit{Exp.4} and \textit{Nom.2}. We can expect, therefore, either the strengthening of standing tropical Rossby waves with faster rotation between $P_{rot}=10-4$~days, or we can expect an abrupt transition to another climate state dominated by standing extra tropical Rossby waves. As reported by C15, the maxima of zonal wind speeds decrease abruptly with faster rotations at $L_R/R_P \approx 1$, when standing extra tropical Rossby waves form.

A drop in maximum zonal wind speed with faster rotation can indeed be seen for \textit{Nom.2} and $P_{rot}=5$ days (Figure~\ref{fig: zonalwind}). Closer inspection of flow patterns (not shown) reveals indeed a transition from a climate state with equatorial superrotation to a new climate state with two high latitude wind jets (see C15). \textit{Exp.4}, on the other hand, assumes a climate state dominated by equatorial superrotation for the whole investigated rotation regime, as evidenced by the smooth increase in zonal wind speeds with faster rotations from $P_{rot}=100$ to $P_{rot}=1$~days (Figure~\ref{fig: zonalwind}).

Thus, we conclude that more efficient night side cooling leads via the larger day to night side temperature gradient to a dominance of standing tropical Rossby waves also for small exoplanets with $R_P=1 R_{Earth}$. In contrast to that, weaker thermal forcing (\textit{Nom.2}) allows for the formation of standing extra tropical waves. The emergence of faster equatorial superrotation (tropical Rossby waves) compared to the slower high latitude jets (extra tropical Rossby waves) explains why \textit{Exp.4} has higher zonal wind speeds than \textit{Nom.2} for rotation periods $P_{rot}\leq 6$~days.

\subsubsection{Climate state bifurcation II: $L_R/R_P \leq 0.5$}
For even faster rotations, the extra tropical Rossby wave has another possible transition, when $L_R/R_P\leq 0.5$. This is at $P_{rot}=3$~days for \textit{Exp.4}. In this rotation regime, the tropical and extra tropical Rossby wave can both fit on the planet allowing for ’mixed states’ with combined tropical and extra tropical Rossby waves. Indeed, mixed states can be found for \textit{Nom.2} in the expected rotation regime (Figure~\ref{fig: zonalwind}, arrows indicating 'mixed states'). For $P_{rot}$= 1 days, however, \textit{Nom.2} goes again into full equatorial superrotation which can be seen by the very fast zonal wind speeds of $u=275$~m/s that the model assumes (Figure~\ref{fig: zonalwind} at $P_{rot}=1$~day). Different possible climate states for the fast rotation regime also explain why surface temperatures, in particular at the night side, can vary so strongly between different models with rotations faster than $P_{rot}= 10$~days (Figure~\ref{fig: surf_ave_Rp}).

We concluded in C15 that the tropical Rossby wave is stronger in \textit{Nom.2} than in the model used by \cite{Edson2011}, where extra tropical Rossby waves dominate even for very fast rotations $P_{rot}=1$~day. We find that the model with more efficient night side cooling (\textit{Exp.4}) is even more strongly dominated by the tropical Rossby wave in the short but also in the long period rotation regime. Stronger thermal forcing via efficient night side cooling rather favours formation of standing tropical Rossby waves. This result is not entirely unexpected, because one of the driving mechanisms for superrotation is longitudinal heating differences in the form of the day side to night side temperature gradient that trigger the formation of Kelvin waves \citep{Showman2011}.

This point is still worthwhile emphasizing, because the comparison between our nominal model with the model of \cite{Edson2011} yielded the comparatively warm night side surface temperatures as one possible source of deviation. One may suspect, therefore, that stronger thermal forcing may lead to climate states dominated by extra tropical Rossby waves as demonstrated by the model of \cite{Edson2011}. \textit{Exp.4} has shown that stronger thermal forcing alone has in fact the opposite effect.

\subsubsection{Circulation}
\label{sec: circ_ns}

\begin{figure}
\includegraphics[width=0.48\textwidth]{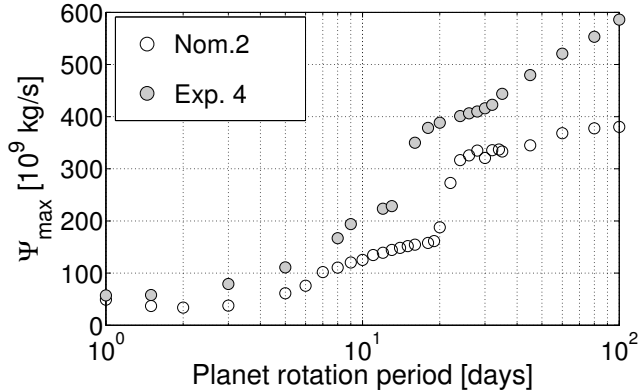}
\caption{Maxima of the meridional mass transport stream function $\Psi$ on the Northern hemisphere for the direct circulation cell in \textit{Nom.2} and \textit{Exp.4}, both for $R_P=1 R_{Earth}$.}
\label{fig: circulation}
\end{figure}

The larger day side to night side and thus also pole to equator surface temperature gradient at the day side\footnote{We can use day side to night side temperature differences to infer pole to equator differences, because the surface temperatures at the night side are relatively uniform compared to the day side: At the night side, the surface temperature gradient is only about $\Delta T \leq 20$~K. On the day side, the surface temperature contrast is $\Delta T \geq 60$~K. See also C14 (their Figure~4) or \cite{Merlis2010} (their Figure~1). Therefore, we can assume to first order that the terminators and the poles are as cold as the night side.} in \textit{Exp.4} should promote more efficient circulation for the same planet size and rotation period. In the following, we will investigate how circulation in \textit{Exp.4} changes compared to \textit{Nom.2}.

We identify circulation states by monitoring the maximum and minimum of the meridional mass transport stream function $\Psi$. This is defined as
\begin{equation}
\Psi=\frac{2\pi R_P}{g}\cos\nu \int_0^p \bar{v} dp',\label{eq: Psi}
\end{equation}
where $\nu$ is latitude and $\bar{v}$ is the zonal and temporal mean of the meridional velocity component $v$ at a given latitude\footnote{Note that the horizontal wind velocity is $\vec{v}=(u,v)$, where $u$ is the zonal and $v$ the meridional component.}. $\Psi$ is positive for clockwise circulation and negative for counter-clockwise circulation. Thus, the direct circulation cell has positive $\Psi$ and the secondary cell negative $\Psi$ on the Northern hemisphere.

The overall circulation efficiency of the direct equatorial cell is indeed greater for \textit{Exp.4} compared to the nominal model (See also C15, Figure 17). The slow rotation regime is, thus, generally dominated by two large direct circulation cells (\textit{state 0}) - as expected from previous studies like \cite{Edson2011} and \cite{Navarra2002}. When \textit{Exp.4} reaches the $\lambda_R/R_P \leq 1$ transition with faster rotation, the circulation evolves into a state with an embedded reverse circulation cell (\textit{state 1}). The development coincides with a climate state dominated by standing tropical Rossby waves at $P_{rot} =$ 18-28 days (Section~\ref{sec: ns_lambda1}). In Figure~\ref{fig: circulation}, upper panel, the development of a \emph{state 1} circulation can be inferred by the steep drop in direct circulation cell strength with faster rotation at $P_{rot}=18$~days. The nominal model (\emph{Nom.2}) forms \textit{state 1} for slower rotations at $P_{rot}=22$~days.

We attribute the formation of \textit{state 1} circulation for faster planet rotation in \textit{Exp.4} compared to \textit{Nom.2} to the presence of weak Rossby waves that already appear for slow rotations $P_{rot}=100$~days in \textit{Exp.4} (Section~\ref{sec: ns_slow}). While it was found that the Rossby waves play only a minor role in the slow rotation regime due to their weakness, they are apparently still strong enough to hamper the formation of embedded reverse circulation cells in the intermediate rotation regime. Also the transition to circulation states with two fully vertical extended circulation cells (\textit{state 2}) is shifted towards faster rotations in \textit{Exp.4} compared to \textit{Nom.2} (Table~\ref{tab: circulation}).

\begin{table}
\caption{Circulation cells.}
\begin{tabular}{c|c|c|c|c}
\hline
 & state 0 &state 1 & state 2 & state 3\\
\hline
$R_P$ & $P_{rot}$&$P_{rot}$ & $P_{rot}$ & $P_{rot}$\\
{[}$R_{Earth}$]&[days] & [days] & [days]  & [days]\\
\hline
Nom. 2 &100 - 22& 20 - 13& 12 - 1.5& 1 \\
Exp. 4 &100 - 18${}^a$ & 17 - 8& 5 - 3& 1.5 -1\\
\hline
\end{tabular}
\newline
${}^a$ where \textit{state 0} contains in this case a weak polar cell
\label{tab: circulation}
\end{table}

\section{Surface friction and planetary boundary layer variation}

In the following, we will investigate how climate dynamics changes for variations of the extent of planetary boundary layer (PBL) and the efficiency of surface friction. For these experiments, we use the nominal model (\emph{Nom.1}) for a Super-Earth planet with $R_P = 1.45 R_{Earth}$ and terrestrial bulk density, as introduced in C14, as a basis. See Table~\ref{tab: ns_models} for the relevant parameters of \textit{Nom.1}.

\subsection{Rayleigh friction prescription}

The Rayleigh friction mechanism prescribed in our model is:
\begin{equation}
\mathcal{\vec{F}}_v=-\frac{1}{t_{fric}}\vec{v},
\end{equation}
where $t_{fric}$ is defined as
\begin{equation}
t_{fric}=t_{s,fric} \max\left(0,\frac{\frac{p}{p_s}-\frac{p_{PBL}}{p_s}}{1-\frac{p_{PBL}}{p_s}}\right),
\end{equation}
where $p_{PBL}$ is the pressure at the upper vertical limit of the planetary boundary layer (PBL) and $t_{s,fric}$ is the maximum surface friction. For the Earth, $t_{s,fric}=1$~day and $p_{PBL}=700$~mbar is assumed \citep{HeldSuarez1994}. But even for Mars, several models assume $t_{s,fric}$ and $p_{PBL}$ that vary by orders of magnitude (see C15 for a more detailed discussion). \cite{HengVogt2011} studied variations of $t_{s,fric}$ with four experiments, but did not vary the extent of the PBL and covered only one rotation period: $P_{rot}=36.5$~days. Furthermore, they changed at the same time radiative forcing and only reported surface temperatures and velocities.

In this study, we use \textit{Nom.1} as a basis and keep $t_{rad}$ the same for every experiment in this subsection (Table~\ref{tab: ns_models}). We study surface friction time scales for $t_{s,fric}=0.1,1,10$ and $100$~days and assign for $t_{s,fric}=1$~days three upper extents of the planet boundary layer: $p_{PBL}=700,800$ and $900$~mbar (see Table~\ref{tab: fric_models}). These variations cover the values that were identified in C14 from climate models of Solar System terrestrial planets. For each scenario, we cover the whole relevant rotation period range $P_{rot}=1-100$~days to monitor coherently climate dynamics transitions at $\lambda_R/R_P=1$, $\lambda_R/R_P=0.5$, $L_R/R_P=1$ and $L_R/R_P=0.5$ due to surface boundary variations.

\begin{table}
\caption{Parameters of the experiments used to investigate different surface prescriptions}
\begin{tabular}{l|c|c|c|c|}
\hline
parameter & Exp. 5 & Nom. 1 & Exp. 6 & Exp. 7 \\
\hline
 $t_{s, fric}$ [days] & 0.1 & 1& 10 &100 \\
 $p_{PBL}$ [mbar] & 700 & 700 & 700 & 700 \\
\hline
parameter & Exp. 8 & Exp. 9 & Nom. 1 \\
\hline
 $t_{s, fric}$ [days] & 1& 1 & 1\\
 $p_{PBL}$ [mbar] & 800 & 900 & 700\\
\hline
\end{tabular}
\label{tab: fric_models}
\end{table}

\begin{figure}
\includegraphics[width=0.48\textwidth]{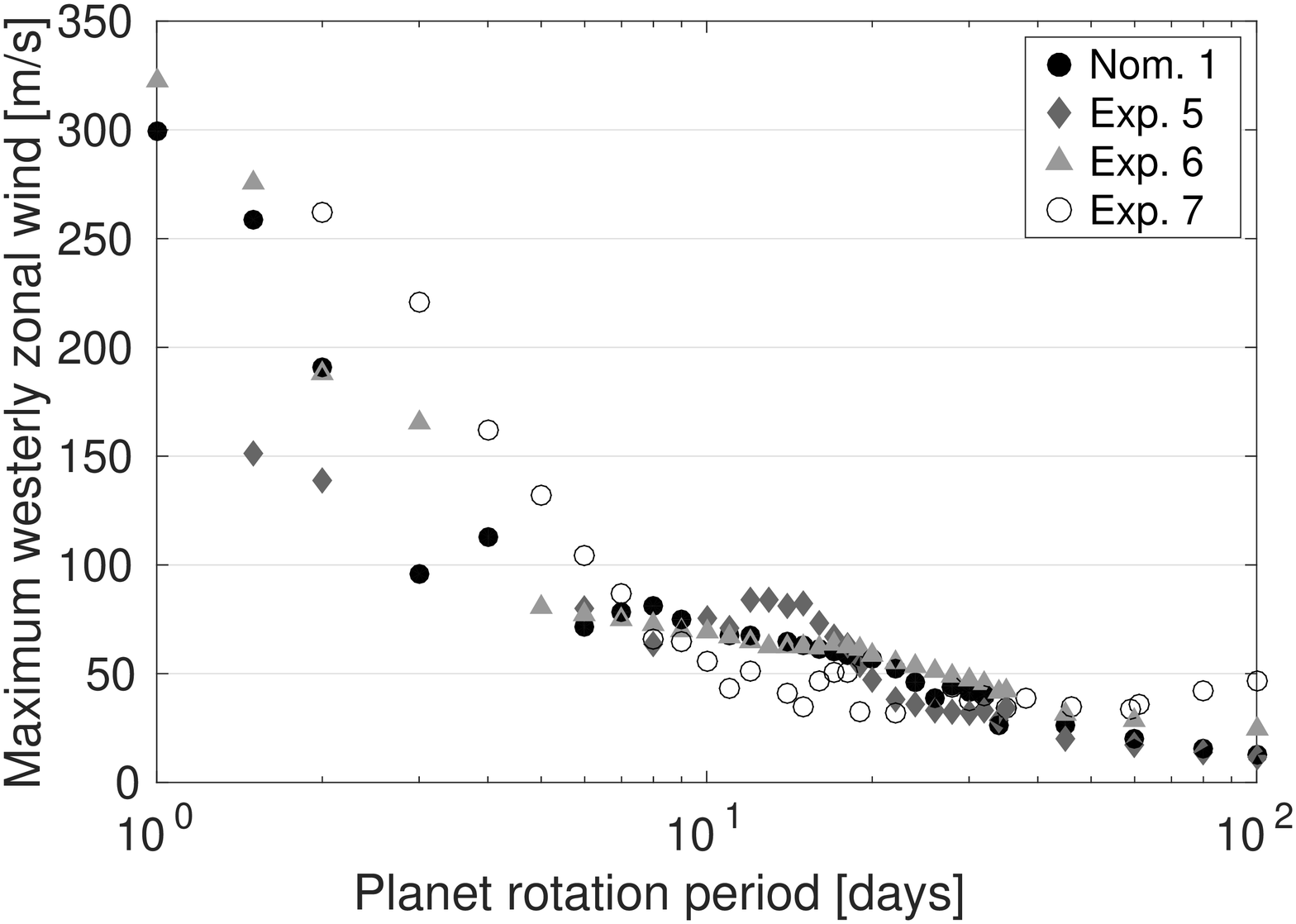}
\includegraphics[width=0.48\textwidth]{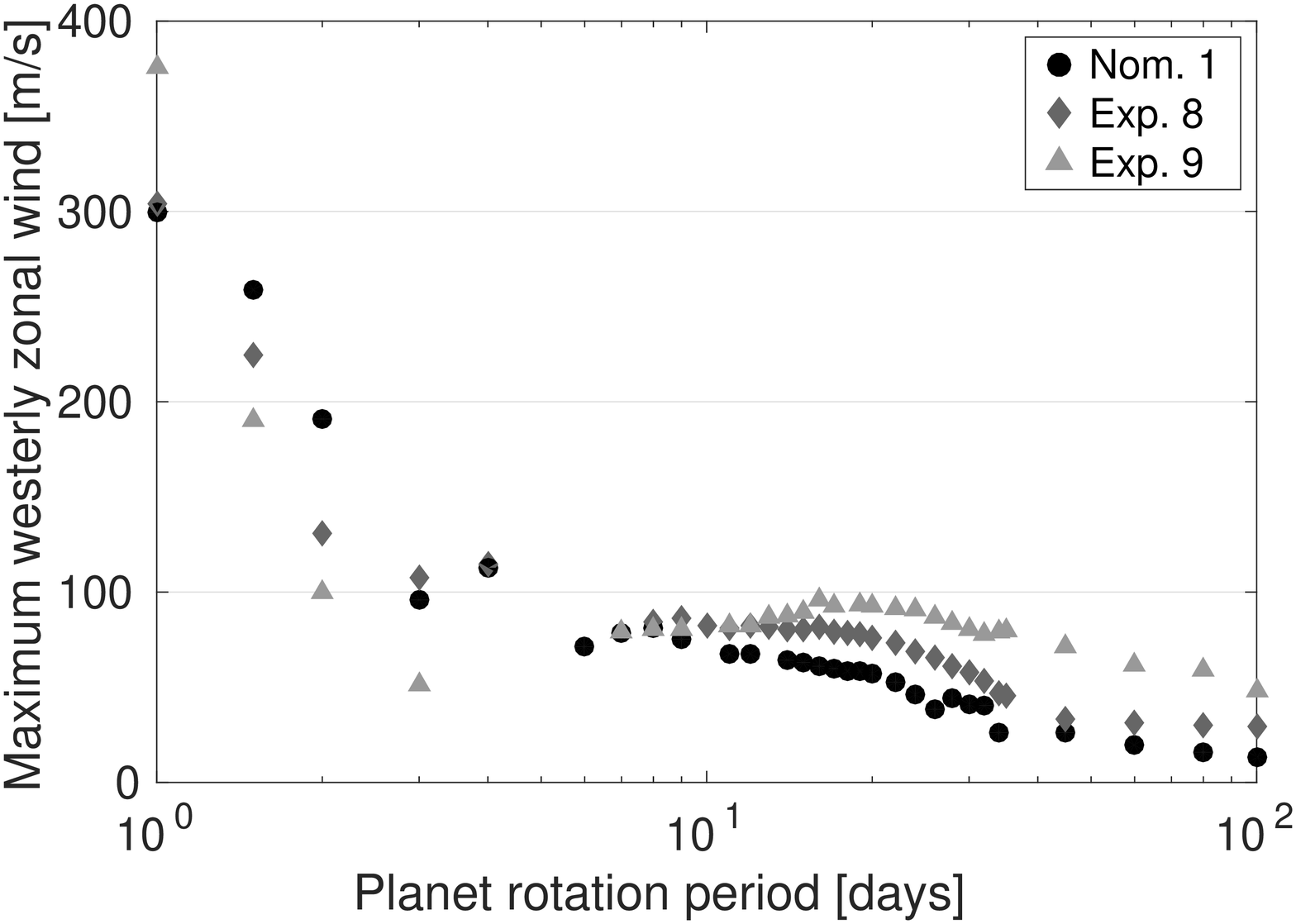}
\caption{Maximum of westerly wind speeds in m/s for different surface friction time scales $t_{s,fric}=0.1-100$~days (top) and different PBL extent $p_{PBL}=700-900$~mbar (bottom).}
\label{fig: zonal_wind1}
\end{figure}

\subsection{Effect of surface friction and PBL extent on large scale dynamics}

To roughly identify climate state changes, we monitor again zonal wind speeds, which are defined as positive for westerly winds and negative for easterly winds.

\subsubsection{Zonal wind}

\begin{figure}
\includegraphics[width=0.48\textwidth]{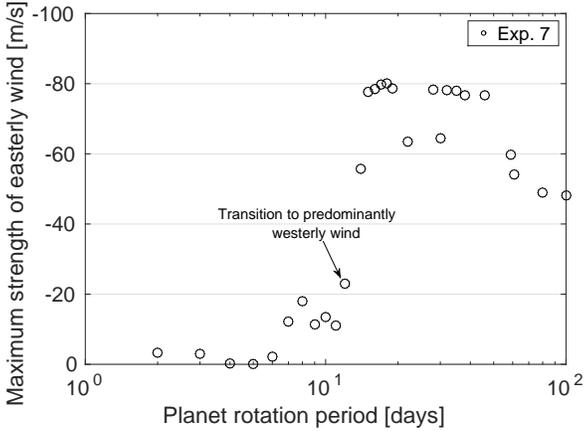}
\caption{Maximum strength of easterly wind speeds in m/s for surface friction time scale $t_{s,fric}=100$~days and $R_P=1.45 R_{Earth}$.}
\label{fig: zonal_wind2}
\end{figure}

Figure~\ref{fig: zonal_wind1}, upper panel, shows that westerly wind speeds generally increase as surface friction efficiency decreases for fast rotation ($P_{rot}\leq 10$~days) and slow rotations ($P_{rot}\geq 40$~days). For intermediate rotations ($P_{rot}=10-40$~days), zonal wind speed evolution for different surface friction time scales is surprisingly complex. For example, we find for $P_{rot} = 10-20$~days that the model with the most efficient surface friction (\textit{Exp. 5}) has strongest westerly winds. The model with the least efficient surface friction (\emph{Exp.7}) has in contrast to that the weakest westerly winds.

Even more surprisingly, Figure~\ref{fig: zonal_wind2} reveals that \textit{Exp.7} shows predominantly easterly zonal winds in the slow and intermediate rotation regime ($P_{rot}$ = 12 -- 100 days) with relatively high zonal wind speeds for the slow rotation regime: $u=-50$ to $-80$~m/s. As rotation period decreases, \textit{Exp.7} transits into a regime with westerly zonal wind tendency at $P_{rot}$ = 12 days (Figure~\ref{fig: zonal_wind2}). The transition occurs well in the standing tropical Rossby wave rotation regime ($\lambda_R/R_P <$ 1) (See Table~\ref{tab: Rossby wave}).

Models with weak surface friction may shed light on climate dynamics in the intermediate regime between terrestrial planets and Mini-Neptunes without a solid surface. The latter planets should have weak or no lower boundary friction.
Based on the result from this Section, we would assume that Mini-Neptunes generally experience faster wind speeds than terrestrial planets for the same thermal forcing. Indeed, tidally locked Mini-Neptune models for GJ1214b ($P_{rot}=1.58$~days) exhibit fast westerlies with wind speed of up to 2~km/s (e.g. \citet{Menou2012}). GJ1214b experiences, however, sixteen times the stellar irradiation than the terrestrial planets discussed in this work and is thus not in the same thermal forcing regime \citep{Charbonneau2009}.

\begin{figure}
\includegraphics[width=0.48\textwidth]{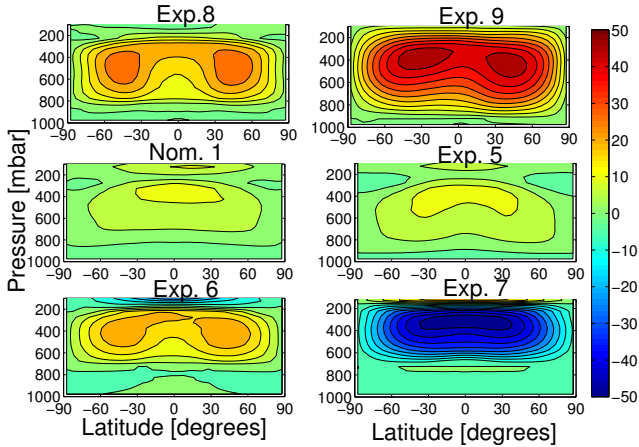}
\caption{Zonal mean of zonal winds in [m/s] for $P_{rot}=100$~days and \textit{Exp.8}, \textit{Exp.9} and \textit{Nom.1}, which have varying different PBL extents ($p_{PBL}$=800, 900~mbar and 700~mbar, respectively) and \textit{Exp.5}, \textit{Exp.6} and \textit{Exp.7}, which have varying surface friction time scales ($t_{s,fric}=0.1, 10$ and 100~days, respectively). Contour levels are 5~m/s.}
\label{fig: Prot100d_zonal_wind}
\end{figure}

The extent of the PBL has also a relatively strong effect on zonal wind speeds as evidenced in Figure~\ref{fig: zonal_wind1}, lower panel. The wind velocities are larger for smaller PBL extent in the intermediate and slow rotation regime ($P_{rot}$ = 10 −- 100 days). This tendency reverses in the fast rotation regime ($P_{rot}$= 1.5 -− 3 days), where the wind speeds are lower for smaller PBL extent. The fast planet rotation regime ($P_{rot}$= 1 -− 3~days) coincides with the Rossby wave regime $L_R/R_P \leq 0.5$ (see Table~\ref{tab: Rossby wave})\footnote{The Rossby radii of deformation are only weakly affected by changes in surface friction and PBL compared to the nominal
model.}, where we found in C15 that standing extra tropical Rossby waves can form alongside standing tropical Rossby waves. In C15, we also showed that climate states with strong extra tropical Rossby waves have lower wind speeds than climate states dominated by tropical Rossby waves that exhibit strong equatorial superrotation. Thus, it appears likely that extra tropical Rossby waves form in \textit{Exp.9} for fast planet rotation periods based on the displayed low wind speeds. The wind speeds of every model with different PBL extent converge towards full strong equatorial superrotation for very fast planet rotation periods, $P_{rot}$= 1~days, and reach high velocities of $u\approx 300$~m/s.

In the following, we will discuss in more detail differences arising from surface boundary variations at climate state transitions with respect to standing tropical and extra tropical Rossby waves as identified in C15. The location of transition regions in rotation period for the set of models investigated in this Section are listed in Table~\ref{tab: Rossby wave}. We compare the results of \emph{Exp.5-9} to the climate states of \textit{Nom.1}, as reported in C15.

\begin{table}
\caption{Rossby wave number transitions for $R_P=1.45 R_{Earth}$.}
\begin{tabular}{c|p{1.2cm}|p{1.2cm}|p{1.2cm}|p{1.2cm}}
\hline
$R_P$ & $P_{rot}$ for $\frac{L_R}{R_P}\approx 0.5$  & $P_{rot}$ for $\frac{L_R}{R_P}\approx 1$ & $P_{rot}$ for $\frac{\lambda_R}{R_P}\approx 1$& $P_{rot}$ for $\frac{\lambda_R}{R_P}\approx 0.5$\\
{[}$R_{Earth}$] & [days] & [days] & [days] & [days]\\
\hline
1.45 & 5 & 10 &34 & 8\\
\hline
\end{tabular}
\label{tab: Rossby wave}
\end{table}

\begin{figure*}
\includegraphics[width=0.8\textwidth]{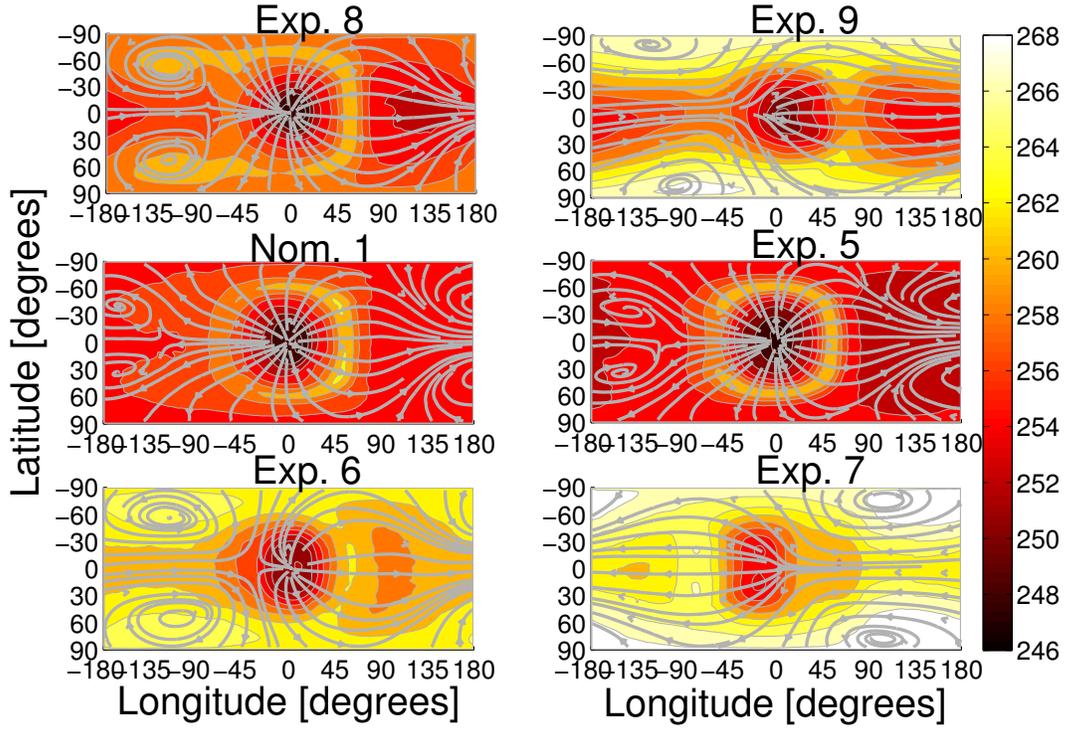}
\caption{ Temperatures (contour interval 2 K) and streamlines of the horizontal flow at p=225 mbar for the $P_{rot}=100$~days simulation of \textit{Experiments 8,9}, the nominal model and \textit{Experiments 5,6,7}.}
\label{fig: Prot100d_p225mbar}
\end{figure*}
\subsubsection{Slow rotations with $\lambda_R/R_P > 1$}
\label{sec: slow_rotPBL}

\begin{figure*}
\includegraphics[width=0.8\textwidth]{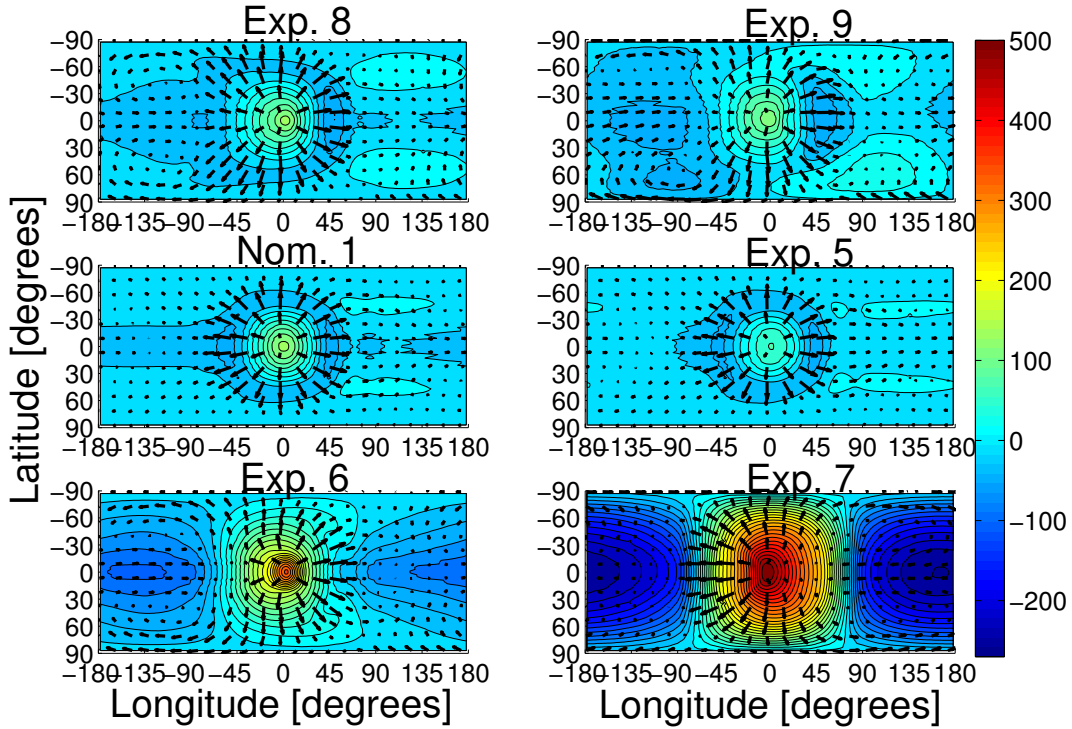}
\caption{Eddy geopotential height in [m] and eddy horizontal wind in [m/s] at p = 225 mbar for the $P_{rot}=100$~days simulation of \textit{Experiments 8,9}, the nominal model and \textit{Experiments 5,6,7}. Contour levels are 20~m and the longest eddy wind vectors are left to right and from top to bottom 25.0, 34.4, 39.4, 37.8, 45.0 and 47.0~m/s, respectively.}
\label{fig: Prot100d_eddy}
\end{figure*}

As already shown in the previous Section, \textit{Exp.7} with very weak surface friction ($t_{s, fric}$ = 100 days) has predominantly easterly winds in  the slow planet rotation regime ($P_{rot} \geq 34$~days). This behaviour is in contrast to all other experiments (\emph{Exp.5}, \textit{Exp.6}, \textit{Exp.8} and\textit{ Exp.9}) and the nominal model (\textit{Nom.1}). This is evident in the zonal mean of the flow (Figure~\ref{fig: Prot100d_zonal_wind}, bottom-right) and even in the horizontal flow pattern at $p=225$~mbar (Figure~\ref{fig: Prot100d_p225mbar}, bottom-right).

Inspection of the eddy geopotential height and wind reveals very deep negative geopotential height perturbations that are centred at the equator and located east and west of the substellar point between $\phi=-90^{\circ}$ to $\phi=-180^{\circ}$ and between $\phi=+90^{\circ}$ to $\phi=+180^{\circ}$ (Figure~\ref{fig: Prot100d_eddy}, bottom right). Furthermore, there are strong easterly eddy winds $\vec{v}'$ that are confined along the equator between $\phi=-90^{\circ} - 0^{\circ}$. Interestingly, again two cyclones are present in the horizontal flow in the upper atmosphere. The vortices are, however, 'flipped' with respect to their counterparts with 'normal' westerly winds in the same rotation regime but with stronger surface friction (\emph{Exp.5}, \textit{Exp.6}, \textit{Exp.8} and \emph{Nom.1}) from $\phi=-135^{\circ}$ to $\phi=+135^{\circ}$ (Figure~\ref{fig: Prot100d_p225mbar}).

The zonal confinement of eddy winds along the equator and deep geopotential anomalies centered over the equator in \emph{Exp.7} are  indications of a Kelvin wave \citep{Matsuno1966} that may be triggered by a very cold night side and thus a strong horizontal temperature gradient. However, Kelvin waves only propagate eastward, which is the reverse direction of the observed zonal winds. The cyclones normally indicate the presence of a large scale Rossby wave that should propagate in the westerly direction. The shear between Kelvin and Rossby wave should thus lead to a westerly equatorial wind not an easterly \citep{Showman2011}. In addition, the slow planet rotation ($P_{rot}=100$~days) is too slow to elicit a strong Rossby wave. Thus, we are currently at a loss to explain this peculiar easterly flow in \emph{Exp.7}.

In general, we find that the depth of negative geopotential perturbations in the slow rotation regime is larger with less efficient surface friction as comparison in particular between \textit{Exp. 6}, \textit{Exp.7} and \textit{Nom.1} shows (Figure~\ref{fig: Prot100d_eddy}, bottom and middle-left panel). The deeper negative geopotential perturbations suggest that the night side surfaces are comparatively cold in these models. The negative geopotential perturbations $z'$ are connected to cyclones that become stronger with decreasing $z'$. The stronger vortices result in turn in warmer temperature maxima in the upper atmosphere (Figure~\ref{fig: Prot100d_p225mbar} bottom and middle-left panel) and also in faster zonal wind speeds $|u|$ (Figures~\ref{fig: zonal_wind1}, upper panel, and \ref{fig: zonal_wind2}).

Models with more compressed PBL extent (\textit{Exp.8} and \textit{Exp.9}) show signs of additional dynamics apart from direct circulation in the slow planet rotation regime (Figure~\ref{fig: Prot100d_eddy}, top panels). Direct circulation is visible in the eddy geopotential height field as a strong positive geopotential height anomaly over the substellar point that indicates the upwelling branch of direct circulation. \textit{Exp.8} and \textit{Exp.9} show that negative geopotential height anomalies appear along with weak tropical Rossby wave-like gyres at mid-latitudes $\nu=\pm 60^{\circ}$ and $\phi=-135^{\circ}$. These Rossby gyres are particulary evident in \textit{Exp.9} (Figure~\ref{fig: Prot100d_eddy}, top-right).

The appearance of these additional dynamical features is surprising, because the $P_{rot}=100$~days simulations are in a Rossby wave number regime with $\lambda_R /R_P>1$, more precisely: $\lambda_R /R_P =$1.75. In the nominal model, standing tropical Rossby wave features were only found for $\lambda_R /R_P \approx 1$  (C15). For larger $\lambda_R /R_P$ the Rossby wave should not be able to 'fit' on the planet.

However, already \textit{Exp.4} with efficient night side cooling has demonstrated that weak tropical Rossby wave-like gyres can form already for $P_{rot}=100$~days (Section~\ref{sec: ns_slow}). It was further shown in Section~\ref{sec: ns_lambda1} that \emph{Exp.4} only assumes a climate state dominated by standing tropical Rossby, if $\lambda_R /R_P<1$ - as expected from C15. The evidence of Rossby waves in \emph{Exp.8} and \emph{Exp.9} confirms our previous statement that Rossby wave gyres can already form for $\lambda_R /R_P$ not much larger than unity. In Section~\ref{sec: ns_slow}, we associated the particularly strong thermal forcing due to an increase in night side cooling efficiency with a stronger preference of tropical Rossby waves. \textit{Exp.8} and \textit{Exp.9} show deep negative geopotential perturbations over the night side that are deeper than similar features in \textit{Exp.4}. These perturbations indicate particularly cold night side surfaces and thus stronger thermal forcing compared to the nominal model (\textit{Nom.2}) but also to \textit{Exp.4}. The triggered Rossby wave is therefore stronger in the slow rotation regime of \textit{Exp.8} and \textit{Exp.9} compared to \textit{Exp.4}. The horizontal flow patterns at $p=225$~mbar already show equatorial jets instead of divergent flow over the substellar point for \emph{Exp.9} (Figure~\ref{fig: Prot100d_p225mbar}). The abnormal climate state with easterly flow in \textit{Exp.7} may be triggered by even stronger thermal forcing due to a particularly cold night side compared to \textit{Exp.8} and \textit{Exp.9}.

Thus, we conclude that cold night sides can trigger very strong thermal forcing between the day side and the night side that leads to climate states with either strong westerly or easterly flow. This flow can already overcome the 'normal' divergent flow on the top of the atmosphere for slow planet rotation ($P_{rot}=100$~days) and $\lambda_R /R_P =$1.75.
\begin{figure*}
\includegraphics[width=0.8\textwidth]{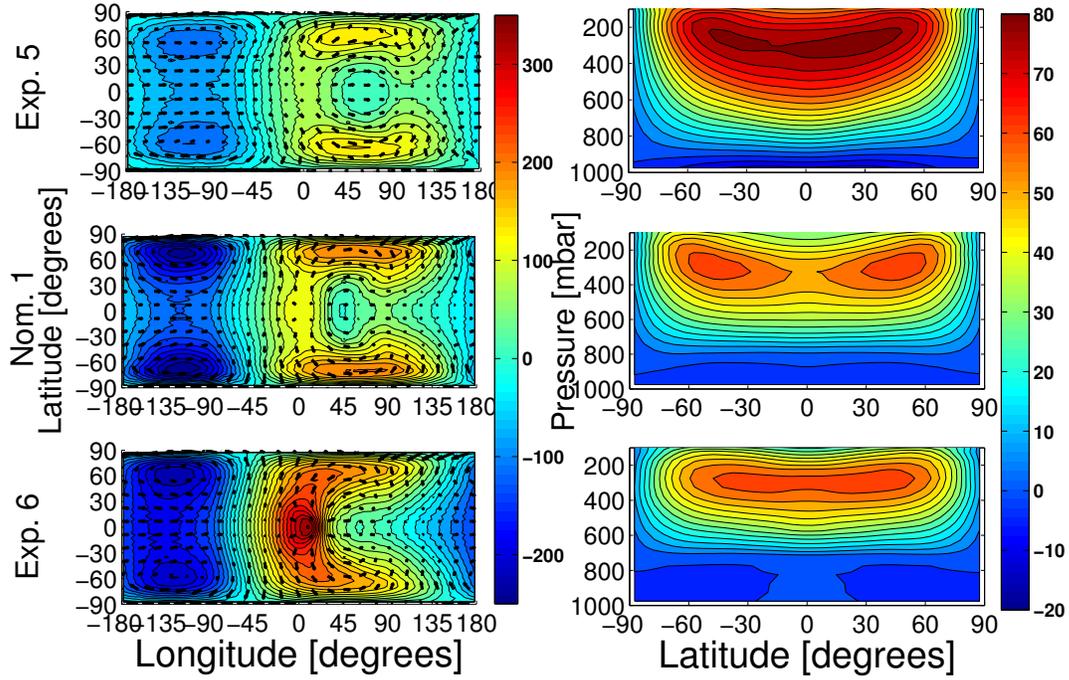}
\caption{Left: Eddy geopotential height in [m] and eddy horizontal wind in [m/s] at p = 225 mbar. Right: Zonal mean of zonal wind in [m/s]. Both properties for the $P_{rot}=15$~days simulation of \textit{Exp. 5, Nom.1,} and \textit{Exp. 6} from top to bottom. Contour levels of eddy geopotential heights are 20~m and the longest eddy wind vectors are from top to bottom 30.7, 47.0 and 44.3~m/s, respectively.}
\label{fig: Prot15d}
\end{figure*}

\subsubsection{Intermediate rotations with $\lambda_R/R_P \leqslant 1$}
\label{sec: fric_lambda1}
\begin{figure*}
\includegraphics[width=0.8\textwidth]{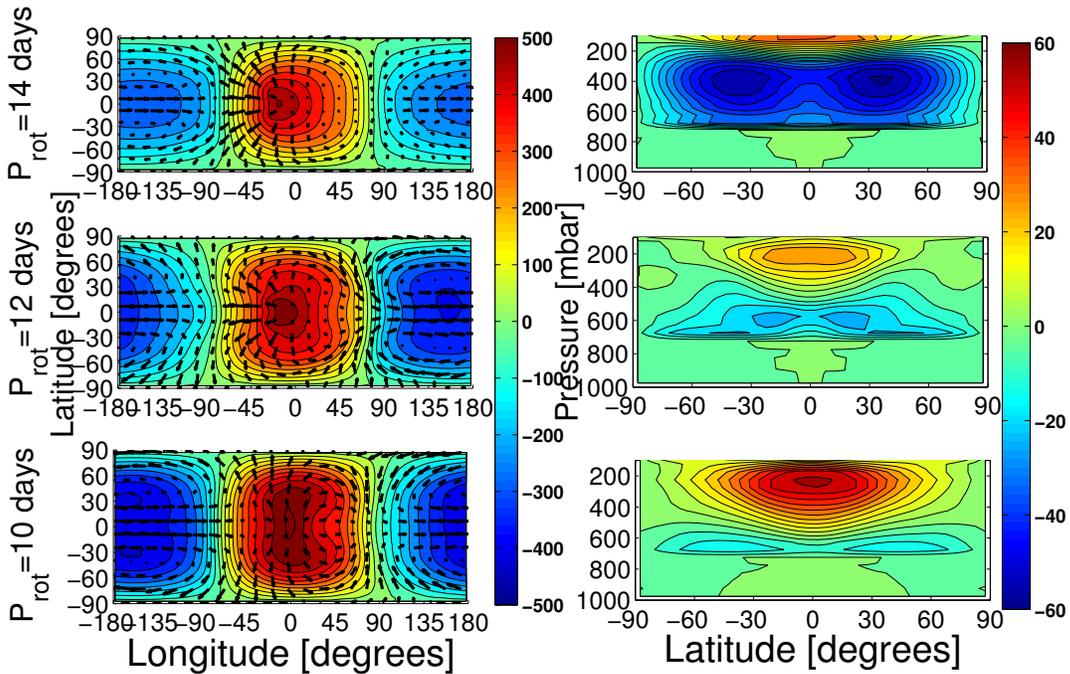}
\caption{Left: Eddy geopotential height in [m] and eddy horizontal wind in [m/s] at p = 225 mbar. Right: Zonal mean of zonal wind in [m/s]. Both properties for \textit{Exp. 7} and for rotation periods $P_{rot}=14,12$ and 10~days from top to bottom. Contour levels of eddy geopotential heights are 50~m and the longest eddy wind vectors are from top to bottom 59.2, 44.7 and 47.5~m/s, respectively.}
\label{fig: taufric100_1}
\end{figure*}

In C15, we found that the intermediate rotation region ($P_{rot} < 34$~days) is associated with the formation of efficient standing tropical Rossby waves. The coupling between the Kelvin and Rossby wave becomes stronger with faster rotation, which leads to an acceleration of zonal winds. There is indeed a rise in zonal wind speeds with faster rotation between $P_{rot}=10-35$~days for all models except \emph{Exp.7} (Figure~\ref{fig: zonal_wind1}). This indicates that the Rossby wave increases in strength with faster rotation periods also for \textit{Exp.8} and \textit{Exp.9} that were shown in the previous Section to already have strong Rossby waves for relatively slow planet rotations.

The model with very efficient surface friction $t_{s,fric}=0.1$~days (\emph{Exp.5}) experiences the strongest zonal wind acceleration with faster rotation in the intermediate rotation regime. Full acceleration is established at rotation period $P_{rot}$= 15 days and continues with faster rotation until a new rotation regime is entered with $L_R/R_P\leq 1$ at $P_{rot}=10$~days (Figure~\ref{fig: zonal_wind1}, upper panel). Apparently, the coupling between the standing Rossby waves and the Kelvin wave is more efficient with stronger surface friction in the intermediate rotation regime. This strong coupling can also be inferred from the eddy geopotential height field at $P_{rot}$ = 15 days (Figure~\ref{fig: Prot15d}, top-left), which shows in contrast to \textit{Exp.1} and \textit{Nom.1} (Figure~\ref{fig: Prot15d}, middle and bottom left), a very weak ($z'\approx 75$~m) positive geopotential height anomaly over the substellar point, indicative of only weak upwelling from direct circulation. The fast equatorial superrotation that follows from the efficient Kelvin-tropical Rossby wave is also discernible as a broad strong equatorial westerly jet in the zonal mean of zonal winds of \emph{Exp.5} (Figure~\ref{fig: Prot15d}, top-right).

The model with least efficient surface friction (\emph{Exp.7}) exhibits an easterly jet even in the intermediate rotation regime, that is, for $P_{rot}=100-12$~days. For faster rotations than $P_{rot}=12$~days, the climate state changes abruptly: The eddy geopotential height and wind fields show between $P_{rot}=10 -14$~days the beginning formation of tropical Rossby wave gyres over negative geopotential height perturbations at $\nu=\pm 30^{\circ}$ and $\phi=-160^{\circ}$ that are particularly evident for $P_{rot}$ = 10 days (Figure~\ref{fig: taufric100_1}, left panels). The full formation of standing tropical Rossby waves is accompanied by a switch in the direction of zonal winds from easterly to westerly (Figure~\ref{fig: taufric100_1}, right panels and Figure~\ref{fig: zonal_wind2} for $P_{rot}=10 -14$~days). The standing tropical Rossby waves force \textit{Exp.7} into a 'normal' superrotating state with faster and faster rotation.

\begin{figure}
\includegraphics[width=0.48\textwidth]{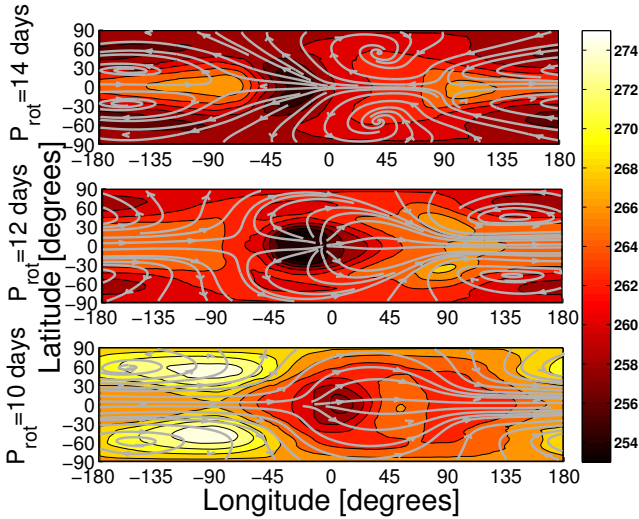}
\caption{Temperatures (contour interval 2 K) and streamlines of the horizontal flow at p=225 mbar for \textit{Exp. 7} and for rotation periods $P_{rot}=14,12$ and 10~days from top to bottom.}
\label{fig: taufric100_2}
\end{figure}

The change in climate dynamics from an easterly towards a westerly flow in \emph{Exp.7} results in drastic changes in the temperatures and horizontal flow patterns at $p = 225$~mbar (Figure~\ref{fig: taufric100_2}). With establishing a climate state dominated by tropical Rossby waves and equatorial superrotation with $P_{rot} \leq 10$~days, the cyclones 'flip' from $\phi=+135^{\circ}$ to $\phi=-135^{\circ}$ and are then associated with local temperature maxima near the west terminator at $\phi=-35^{\circ}$ that are hotter than the temperatures found at $p=225$~mbar for slower rotations (Compare Figure~\ref{fig: taufric100_2} middle and bottom panel). The climate state with predominantly easterly flow has thus a smaller horizontal temperature gradient than the climate state with westerly flow and deep vortices.

Models with reduced PBL extent (\emph{Exp.8} and \emph{Exp.9}) result in comparatively strong standing tropical Rossby wave gyres (not shown) and thus stronger equatorial superrotation as evidenced by the larger zonal wind speeds compared to the nominal model (Figure~\ref{fig: zonal_wind1}, bottom panel between $P_{rot}=10-35$~days). In this respect, the models with reduced surface interface region are similar to \emph{Exp.4} that also showed better coupling between Kelvin and Rossby waves and thus likewise faster zonal wind speeds than the nominal model. This result confirms the conclusion drawn in the previous Section that \emph{Exp.8} and \emph{Exp.9} represent models with particularly strong thermal forcing that induce stronger tropical Rossby waves compared to the nominal model.

\emph{Exp.8} and \emph{Exp.9} never reach, however, the level of strong coupling between Rossby and Kelvin waves as \emph{Exp.5} with the most efficient surface friction. We attribute the stronger zonal winds in \emph{Exp.8}, \emph{Exp.9} and \textit{Exp.4} to the colder night sides and thus larger day to night side temperature gradients compared to the nominal model (\emph{Nom.1}). \textit{Exp.7} shows, however, that the night side should not become too cold, otherwise the climate state may flip into a state with easterly instead of westerly winds. We conclude thus that there are two mechanisms to induce stronger equatorial superrotation: Either by a larger day side to night side temperature gradient or by very efficient surface friction that triggers a stronger coupling between Kelvin and Rossby waves and thus faster equatorial superrotation. The latter appears to only be effective in the intermediate rotation regime.

\subsubsection{Climate state bifurcation I: $L_R/R_P \leq  1$ and $\lambda_R/R_P \leq  0.5$}
\begin{figure*}
\includegraphics[width=0.9\textwidth]{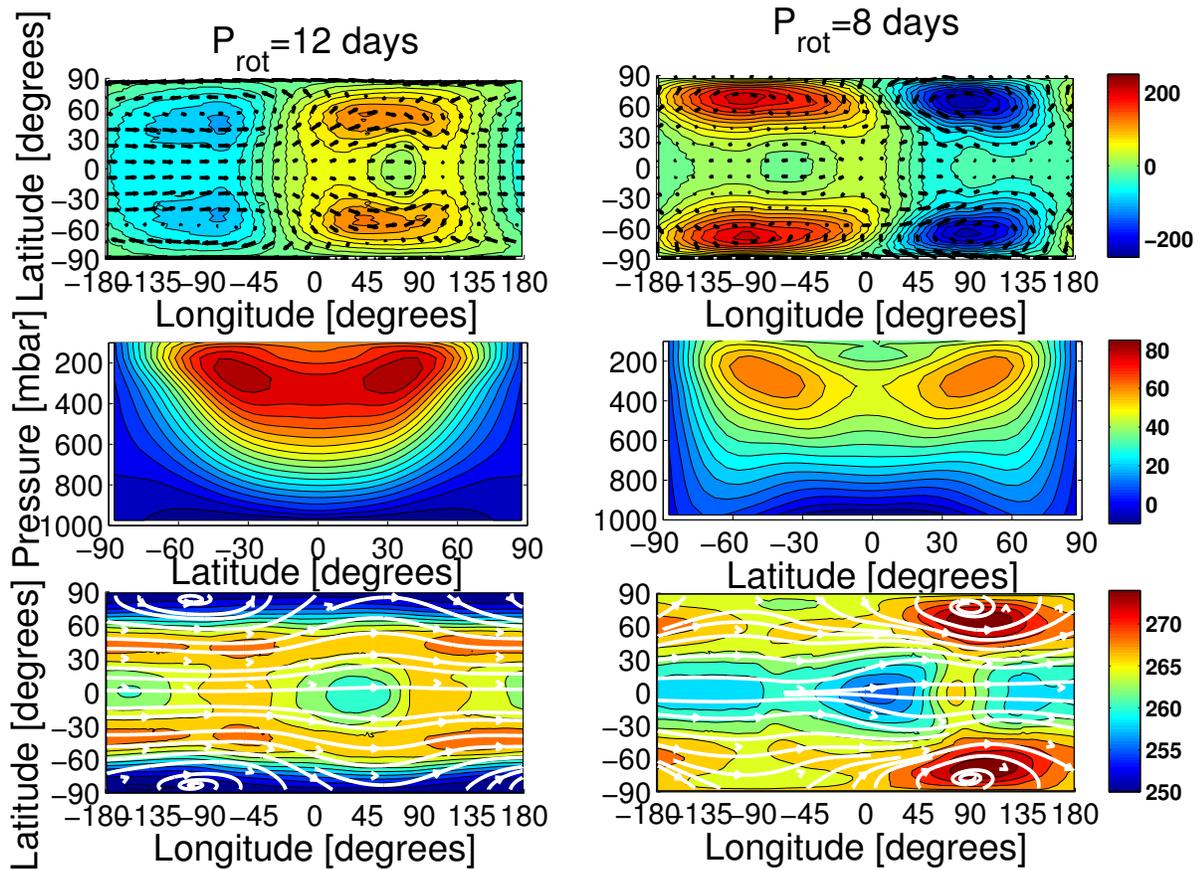}
\caption{Top: Eddy geopotential heights in m (contour interval 20 m) and eddy velocities [m/s]. Middle: Zonal mean of zonal winds in m/s (Contour interval 5~m/s). Bottom: Temperatures (contour interval 2 K) and streamlines of the horizontal flow at p=225 mbar. The $P_{rot}=12$~days (left) and $P_{rot}=8$~days simulation (right) of \textit{Exp. 5} are shown.}
\label{fig: taufric01_LR1}
\end{figure*}

The climate simulations investigated in this Section could, in principle, transit into a new rotation regime for $P_{rot} \leq$~10 days. Because the extra tropical Rossby wave becomes smaller than the planet size  $L_R/R_P \leq 1$ (see Table~\ref{tab: Rossby wave}), the extra tropical Rossby wave can now also form a standing wave. This possible transition almost coincides with the tropical Rossby wave transition with $\lambda_R/R_P \leq 0.5$ at $P_{rot}=8$~days. We concluded in C15 that the climate can thus enter two states in this rotation regime: One dominated by standing extra tropical waves and high latitude jets and one dominated by standing tropical Rossby waves and an equatorial jet, which gains in speed with faster rotation as soon as $\lambda_R/R_P < 0.5$. The formation of a standing extra tropical Rossby wave can already be inferred by an abrupt drop in zonal wind at the $L_R/R_P \leq  1$ transition with faster rotation as shown by \cite{Edson2011} and discussed in C15. Models that remain in a climate state dominated by the tropical Rossby wave experience instead a smooth acceleration of zonal wind speeds with faster rotation around the $\lambda_R/R_P \leq  0.5$ transition as demonstrated in C15.

The abrupt transition to lower zonal wind speeds near $L_R/R_P \leq 1$ can be seen for \emph{Exp. 5} with
$t_{s, fric} = 0.1$~days (Figure~\ref{fig: zonal_wind1} between $P_{rot}=10-12$~days). The drop in zonal wind speeds is accompanied as expected by the formation of two high latitude jets (Figure~\ref{fig: taufric01_LR1}, right-middle panel). The eddy geopotential height field shows at the same time indeed the switch from a climate state dominated by tropical to a state dominated by extra tropical Rossby wave gyres (Compare Figure~\ref{fig: taufric01_LR1} top panel with C15,Figure~2): The extra tropical Rossby wave gyres are located at higher latitudes at $\nu=\pm 70^{\circ}$ compared to $\nu=\pm 45^{\circ}$ in the tropical Rossby wave state (Figure~\ref{fig: taufric01_LR1} top panels). Also the positive and negative geopotential height anomalies are shifted in longitude by about $180^{\circ}$ with respect to each other. The negative geopotential height perturbations are located at $\phi=+90^{\circ}$ in the extra tropical Rossby state and not at $\phi=-110^{\circ}$ as in the tropical Rossby state. The cyclones that are associated with the negative geopotentials also change location accordingly (Figure~\ref{fig: taufric01_LR1}, bottom panels).

The cyclones become again locations of warm temperature maxima after the switch to a climate state dominated by extra tropical Rossby waves. We attribute the formation of temperature maxima to two effects: a) Upwelling is not suppressed by equatorial superrotation. b) The horizontal flow over the substellar point is not confined to the equator and can thus flow into the cyclones to become adiabatically heated (C15).

In C15, we already speculated that efficient surface friction may promote extra tropical Rossby waves. \emph{Exp. 5}, the model with the most efficient surface friction, confirms this hypothesis and shows indeed a climate state dominated by extra tropical waves as soon as $L_R/R_P \leq  1$ with faster planet rotation. All other models remain in climate states dominated by tropical Rossby waves in this rotation regime with predominantly equatorial superrotation with relatively fast zonal wind speeds - at least as long as $L_R/R_P > 0.5$.

\subsubsection{Climate state bifurcation II: $L_R/R_P \leq  0.5$}

For even faster rotations ($P_{rot} \leq 5$~days), another possible climate state transition is possible: We found in C15 that these fast planet rotations allow for three different climate states: One with predominantly equatorial superrotation, one with two high latitude westerly jets and one 'mixed' climate state with features of both previously listed climate states. The latter is possible since both the tropical and extra tropical Rossby wave can simultaneously fit on the planet in this planet rotation period regime; the Rossby radii of deformation of both waves are smaller than half the planet's size. In C15, we found mixed states only for the smallest planet sizes ($R_P\leq 1.25 R_{Earth}$).

In contrast to the results in C15, \emph{Exp. 5} with $R_P=1.45 R_{Earth}$ assumes a climate state with high latitude jets for fast planet rotations between $P_{rot}=10-2$~days. \textit{Exp.5} evolves into a mixed climate state for $P_{rot}$ = 1.5~days with the emergence of an additional equatorial jet (not shown). The existence of climate states with high latitude instead of strong equatorial superrotation can also be inferred from the zonal wind speed evolution of \textit{Exp.5} for $P_{rot}=1.5-2$~days (Figure~\ref{fig: zonal_wind1}, upper panel). Here, \emph{Exp.5} has lower wind speeds compared to other experiments.

We thus conclude that \emph{Exp. 5} is qualitatively the most similar of our models to that of \cite{Edson2011}: \textit{Exp.5} assumes the same climate states not only in the very short ($P_{rot} \leq 5$~days) but also in the short rotation period regime ($P_{rot}\leq 10$~days, see previous Section). The wind speeds of \emph{Exp.5} are still 50\% larger than those reported by \cite{Edson2011}.\emph{Exp. 6} and \emph{Exp.7} with weak surface friction never form standing extra tropical Rossby waves. Efficient surface friction favours the formation of standing extra tropical Rossby wave.

The models with reduced PBL extent, \emph{Exp. 8} and \emph{Exp. 9}, also assume mixed states between $P_{rot}=2-3$~days. These climate states can again be inferred from the relatively low zonal wind speeds, in particular for \textit{Exp.9} (Figure~\ref{fig: zonal_wind1}, lower panel). A compression in PBL extent, keeping otherwise $\tau_{s,fric}=1$~days, is equivalent to raising surface friction efficiency in the very fast rotation period regime.

\subsubsection{Circulation}

\begin{figure}
\includegraphics[width=0.48\textwidth]{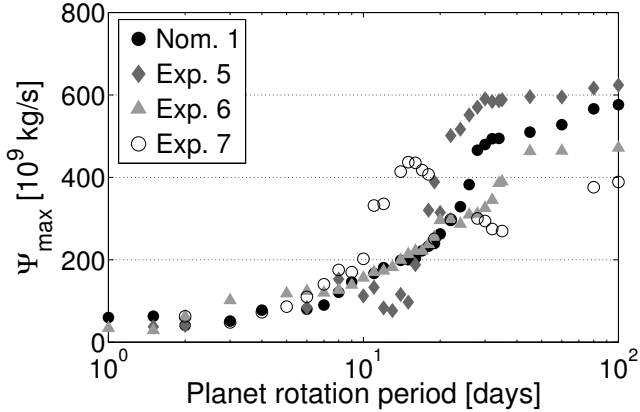}
\caption{Maxima of the meridional mass transport stream function $\Psi$ on the Northern hemisphere for the direct circulation cell for different surface friction time scales and $R_P=1.45 R_{Earth}$ versus rotation period.}
\label{fig: circ_fric}
\end{figure}

\begin{figure}
\includegraphics[width=0.48\textwidth]{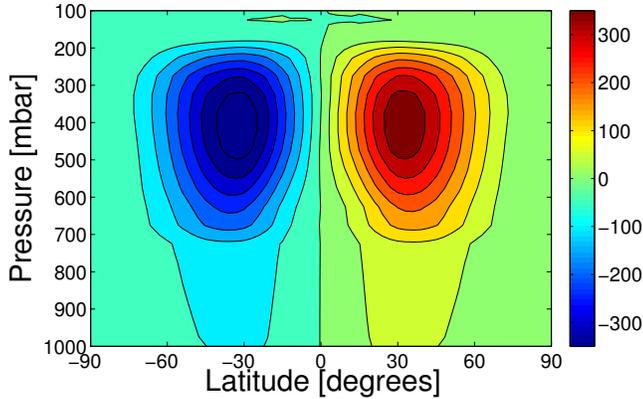}
\caption{ Meridional mass stream function $\Psi$ in units of $10^9$~kg/s for the $P_{rot}=100$~days of \textit{Exp. 7} with $t_{s, fric}=100$~days, where positive values denote circulation in clock-wise and negative values in counter-clockwise direction, respectively.}
\label{fig: Prot100d_Overturn_taufric100}
\end{figure}

Circulation is affected by every aspect of surface boundary variations. Here we monitor again the maxima and minima of the meridional mass stream function $\Psi$ (see Equation~\ref{eq: Psi}) on the Northern hemisphere as introduced in Section~3.4.6 and interpret the results for the expected circulation states: \textit{state 0-3} (see C15, their Figure~17). Generally, models with more efficient surface friction show stronger direct circulation (\textit{state 0}) in the slow rotation regime ($P_{rot}\geq 35$~days), where one large circulation cell fills each hemisphere. Thus, \textit{Exp.5} has the greatest circulation strength $\Psi_{max}$ in the slow rotation regime (Figure~\ref{fig: circ_fric}). Direct circulation in \emph{Exp.7} is not only greatly diminished in strength compared to the other experiments, it also has particularly weak circulation efficiency near the surface as close inspection of $\Psi$ for $P_{rot}=100$~days shows (Figure~\ref{fig: Prot100d_Overturn_taufric100}).

For faster planet rotation, more precisely for $P_{rot}\leq 34$~days, the formation of a new circulation state \emph{state 1} is expected that we associated in C15 with the formation of standing tropical Rossby waves that should form for $\lambda_R/R_P \leq 1$. Circulation \textit{state 1} indeed develops for the expected rotation periods for \textit{Nom.1}, \textit{Exp.5} and \textit{Exp.6}. \textit{State 1} is characterized by an embedded reverse circulation cell inside one large direct circulation cell per hemisphere, when monitoring the meridional mass stream function $\Psi$ for latitude and pressure height (not shown, but see Table~\ref{tab: circulation2} for transition in circulation states based on the form and number of circulation cells in $\Psi$). The beginning formation of a \textit{state 1} circulation can also already be inferred by the steep drop in $\Psi_{max}$ with faster planet rotation for $P_{rot}\leqslant 34$~days in \textit{Nom.1}, \textit{Exp.5} and \textit{Exp.6}. The full formation of the embedded cells, however, are discernible in the meridional stream function $\Psi$ for somewhat faster rotations than $P_{rot}\approx 34$~days, which explains why full establishment of \textit{state 1} is apparent for $P_{rot}\approx 24$~days instead of $P_{rot}\approx 34$~days in \textit{Nom.1}, \textit{Exp.5} and \textit{Exp.6}(Table~\ref{tab: circulation2}). The deviation between Table~\ref{tab: circulation2} and inferred on-set of \textit{state 1} circulation from $\Psi_{max}$ shows that the combined monitoring of the maximum strength of circulation cells ($\Psi_{max}$) and Rossby wave number transitions is necessary to fully interpret changes in the circulation cells in $\Psi$ versus latitude and pressure height. The formation of new weak embedded circulation cells may not readily be apparent in the latter\footnote{See C14 for a more detailed discussion of \emph{state 1}.}.

In contrast to the models with nominal, strong and weak surface friction (\textit{Nom.1}, \textit{Exp.5} and \textit{Exp.6}), the circulation in the model with very inefficient surface friction (\emph{Exp. 7}) appears to gain strength for $P_{rot}\leq 34$~days. $\Psi_{max}$ decreases again with faster rotation for $P_{rot}\leqslant 14$~days (Figure~\ref{fig: circ_fric}). While we cannot explain the anomalous increase in circulation strength with faster planet rotation in \textit{Exp.7}, we can explain the drop in circulation strength for $P_{rot}\leqslant 14$~days. The latter is associated again with the formation of \emph{state 1}-circulation (Table~\ref{tab: circulation2}): It coincides with the switch from a climate state with predominantly easterly winds and suppression of tropical Rossby waves ($P_{rot}>14$~days) to a climate state with fully formed standing tropical Rossby waves and westerly equatorial superrotation ($P_{rot} \leqslant 10$~days). The switch was found in Section~\ref{sec: fric_lambda1} between $P_{rot}=10$ and $14$~days.

Embedded reverse cells are absent in \emph{Exp.7} for intermediate rotation periods ($P_{rot}=10-34$~days), because tropical Rossby waves are suppressed by the abnormal easterly flow. They appear, however, as soon as the tropical Rossby waves fully form. Thus, circulation evolution with rotation period in \textit{Exp.7}, \textit{Nom.1}, \textit{Exp.5} and \textit{Exp.6} confirms that \textit{state 1} circulation is intricately linked to the presence of standing tropical Rossby waves (C15).

\begin{figure}
\includegraphics[width=0.48\textwidth]{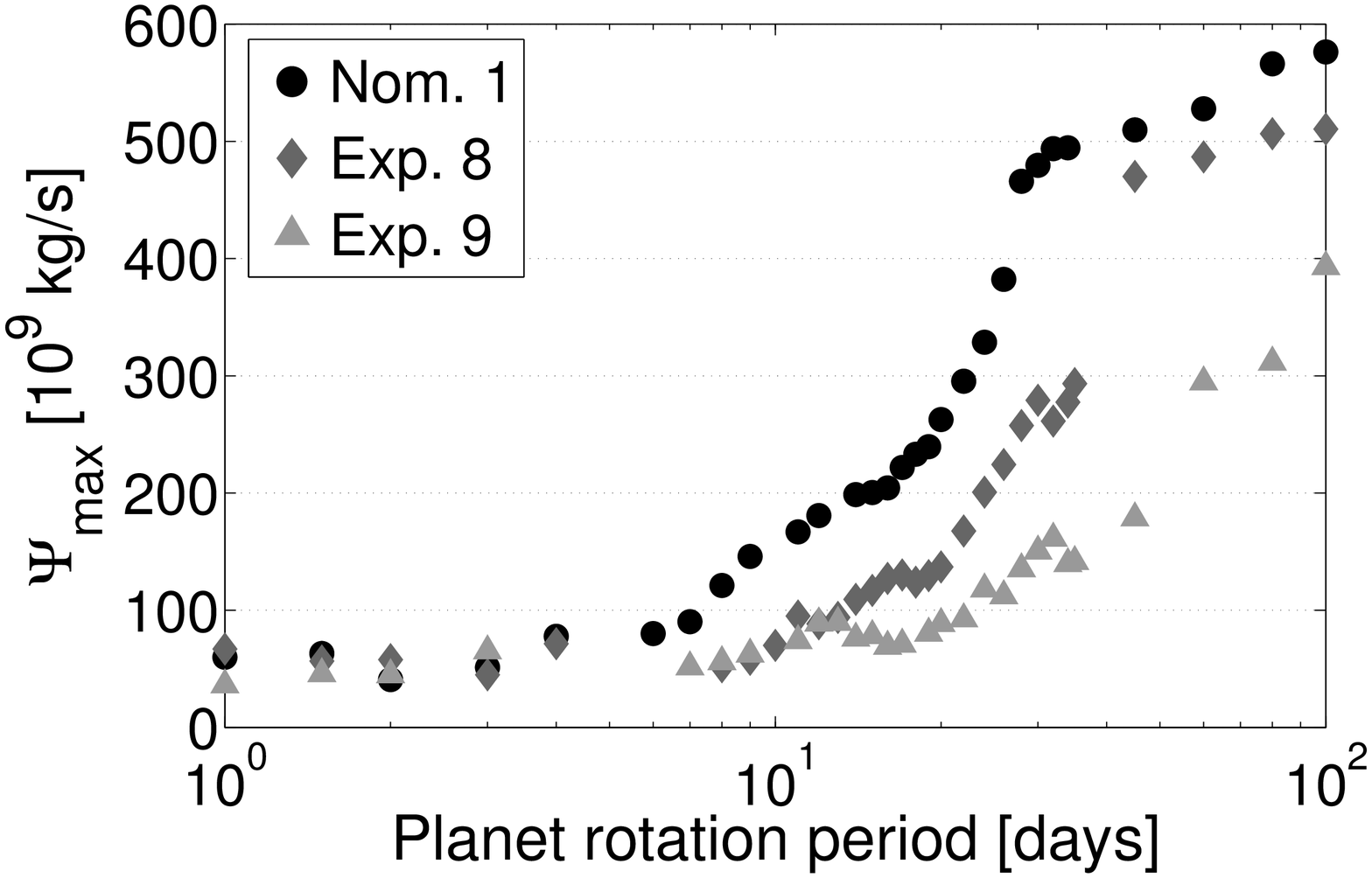}
\includegraphics[width=0.48\textwidth]{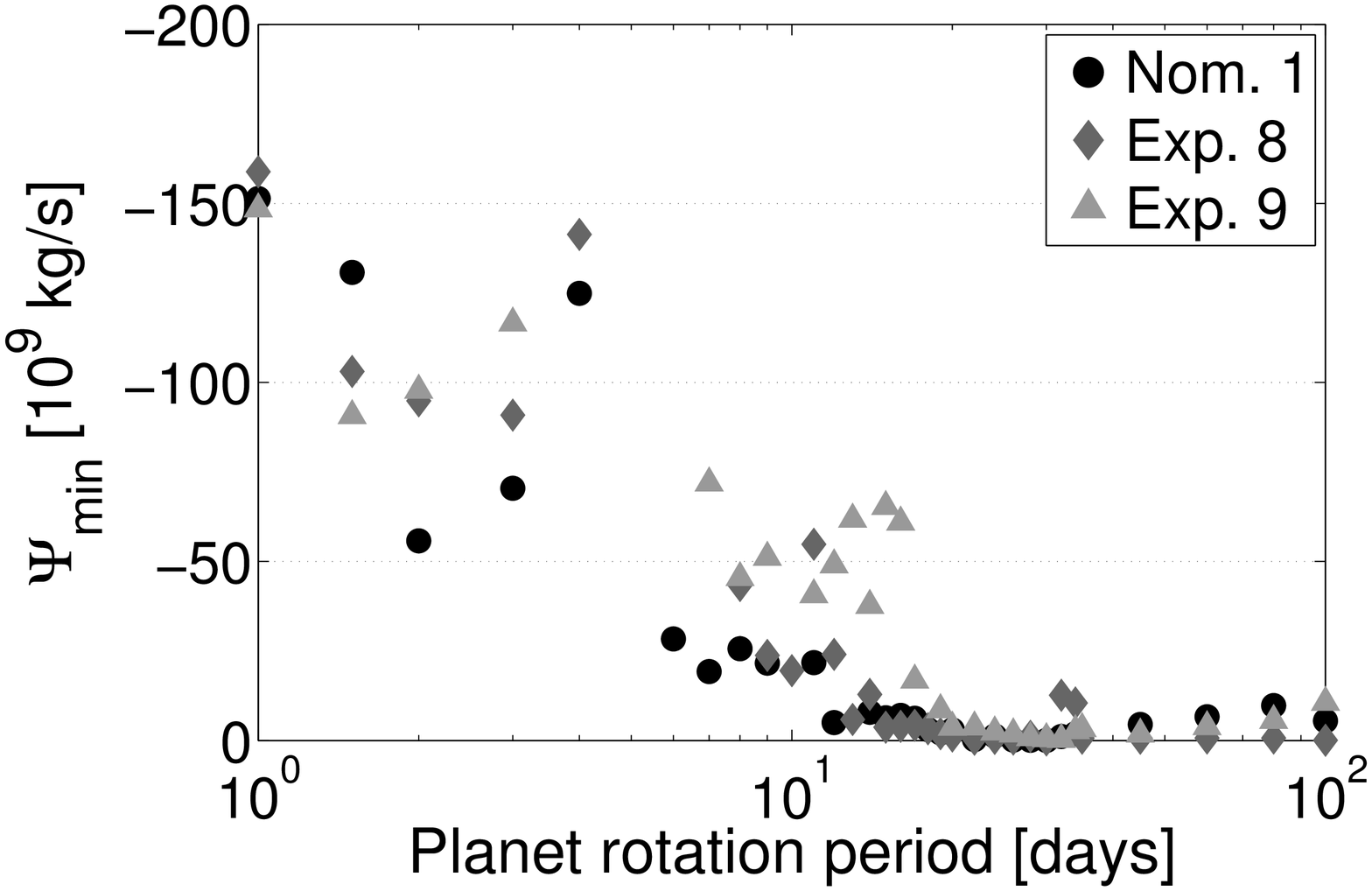}
\caption{ Maxima of the meridional mass transport stream function $\Psi$ on the Northern hemisphere for the direct circulation cell (top) and minima of the secondary circulation cell (bottom) for various PBL extent and $R_P=1.45 R_{Earth}$ versus rotation period.}
\label{fig: circ_PBL}
\end{figure}

\textit{Exp.5} shows in the intermediate rotation regime $P_{rot}=10-15$~days, a particularly strong embedded reverse circulation \textit{state 1}. In Section~4.2.3, we have identified very efficient coupling between tropical Rossby and Kelvin waves for \textit{Exp.5} in this rotation regime. Apparently, this strong coupling leads also to stronger \textit{state 1} circulation. As a result, $\Psi_{max}$ is particularly low for \textit{Exp.5} and $P_{rot}=10-15$~days (Figure~\ref{fig: circ_fric}).

For even faster planet rotations ($P_{rot}\leq 10$~days), also \textit{Nom.1}, \textit{Exp.6} and \textit{Exp.7} show suppression of direct circulation as tropical Rossby waves and thus equatorial superrotation strengthens. In contrast to \textit{Nom.1}, \textit{Exp.6} and \textit{Exp.7}, \emph{Exp. 5} develops from a state with strong equatorial superrotation into a climate state dominated by extra tropical Rossby waves for $P_{rot}\leq 10$~days, as shown in Sections~4.2.4 and 4.2.5. The latter climate state prevents the formation of strong equatorial superrotation and thus maintains direct circulation. Indeed, for $P_{rot} \approx 10$~days, the circulation strength $\Psi_{max}$ increases for \textit{Exp.5} because equatorial superrotation disappears with faster rotation with the switch in climate state (Figure~\ref{fig: circ_fric}). Because direct circulation strength generally decreases with faster rotation, $\Psi_{max}$ does not reach, however, the values for $P_{rot} \geq 30$~days and decreases further in strength with faster rotation.

\begin{table}
\caption{Circulation cells.}
\begin{tabular}{c|c|c|c|c}
\hline
 & state 0 &state 1 & state 2 & state 3\\
\hline
 & $P_{rot}$&$P_{rot}$ & $P_{rot}$ & $P_{rot}$\\
&[days] & [days] & [days]  & [days]\\
\hline
Exp.5 &100 - 22& 20 - 14& 13- 1.5& - \\
Nom.1 & 100 - 26& 24 - 14& 13 - 3 & 2 - 1\\
Exp.6 & 100- 26& 24 - 3 & 2-1${}^a$ & -\\
Exp.7 & 100 - 11& 10 -2 & - & -\\
Exp.8 & 100 - 45& 35- 18& 17 - 2& 1.5-1\\
Exp.9 & 100 & 80 - 20& 22-2& 1.5-1\\
\hline
\end{tabular}
\label{tab: circulation2}
\newline
${}^a$ Vertical break-up of circulation cells.
\end{table}

\emph{Exp. 8} and \emph{9}, the simulations with compressed PBL, show for $P_{rot}=10-100$~days weaker direct circulation compared to the nominal model. The circulation strength is weaker the lower $p_{PBL}$ is (Figure~\ref{fig: circ_PBL}, top panel). The experiments exhibit, furthermore, the appearance of embedded reverse cells (circulation \emph{state 1}) for surprisingly slow planet rotations compared to other models, that is, between $P_{rot}=40-80$~days (Table \ref{tab: circulation2}). The onset of embedded circulation can again already be determined by the steep decline of the strength of the direct circulation cell (Figure~\ref{fig: circ_PBL}, top panel). We have associated in Section~\ref{sec: circ_ns} embedded reverse cells with standing tropical Rossby waves. These form in the nominal model in a rotation period regime for which $\lambda_R/R_P\leq 1$. \emph{Exp.8} and \emph{Exp.9}, on the other hand, have shown to form tropical Rossby waves already at $P_{rot} =$ 100 days for $\lambda_R/R_P=1.75$ (Section~4.2.2). We concluded then that tropical Rossby waves can form under special circumstances -- as presented by \textit{Exp.8.} and \textit{9} -- already for $\lambda_R/R_P$ larger than unity. The Rossby waves in the slow rotation regime are apparently strong enough to induce \textit{state 1} circulation for comparatively slow planet rotators compared to \textit{Nom.1}. This result confirms that embedded reverse cells are inherent properties of climate states with standing tropical Rossby waves - at least in our model prescription.

Models with reduced PBL extent not only develop circulation \emph{state 1} for slow planet rotation, they also develop circulation states with more than one fully vertically extended circulation cell per hemisphere, that is, circulation \emph{state 2} for slower planet rotations compared to the nominal model (\textit{Nom.1}). The formation of circulation \emph{state 2} can be inferred from Table~\ref{tab: circulation2} and Figure~\ref{fig: circ_PBL} (lower panel). The latter shows the substantial strengthening of the secondary circulation cells already for $P_{rot}\approx 22-17$~days. Thus, although direct circulation is suppressed by the tropical Rossby waves, the role of circulation can be taken over by the secondary circulation cells that are not affected by equatorial superrotation. In addition, we note that the strength of the secondary circulation cell is greater for models with compressed PBL (\textit{Exp.8} and \textit{Exp.9}) compared to the nominal model (\textit{Nom.1}).

There are two climate state bifurcations in the short rotation period regime, that is, for $P_{rot}\leq 10$~days. One bifurcation exists around $P_{rot} \approx 10$~days and can lead to climate state transitions for the extra tropical ($L_R/R_P<1$) or tropical Rossby wave ($\lambda_R/R_P<0.5$). The other bifurcation point exists at the extra tropical Rossby wave number $L_R/R_P \approx 0.5$, that is, at $P_{rot} \approx 5$~days. Consequently, there are several possible climate states with different circulation patterns for $P_{rot}\leq 10$~days as evidenced by the strong fluctuation in strength of the secondary circulation cells (Figure~\ref{fig: circ_PBL}, lower panel).

In general, climate states dominated by tropical Rossby waves and thus equatorial superrotation show a vertically compressed direct circulation cell that is often vertically fragmented by embedded reverse circulation (Figure 17 in C15). Circulation \textit{state 2} can, however, lead to a stabilization of circulation cells. Climate states dominated by standing extra tropical Rossby waves have unperturbed direct circulation cells at the equator. \emph{Exp. 5} is in the short rotation regime ($P_{rot} \leq $~3 days) in this state. It will be shown in the following Section that the changes in circulation states due to different climate states also result in different surface temperature evolutions.

\subsubsection{Surface temperatures and habitability}

\begin{figure}
\includegraphics[width=0.48\textwidth]{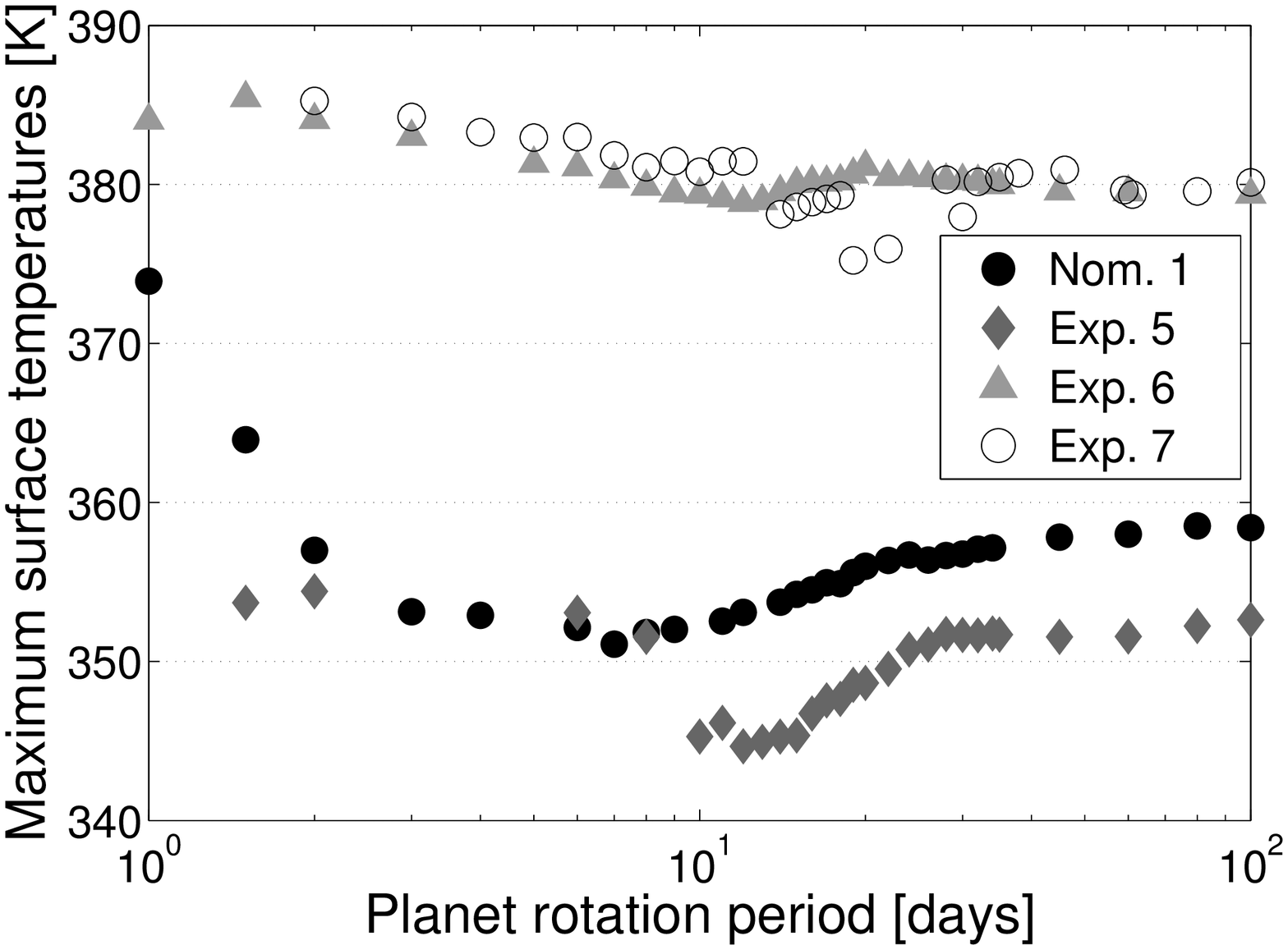}
\includegraphics[width=0.48\textwidth]{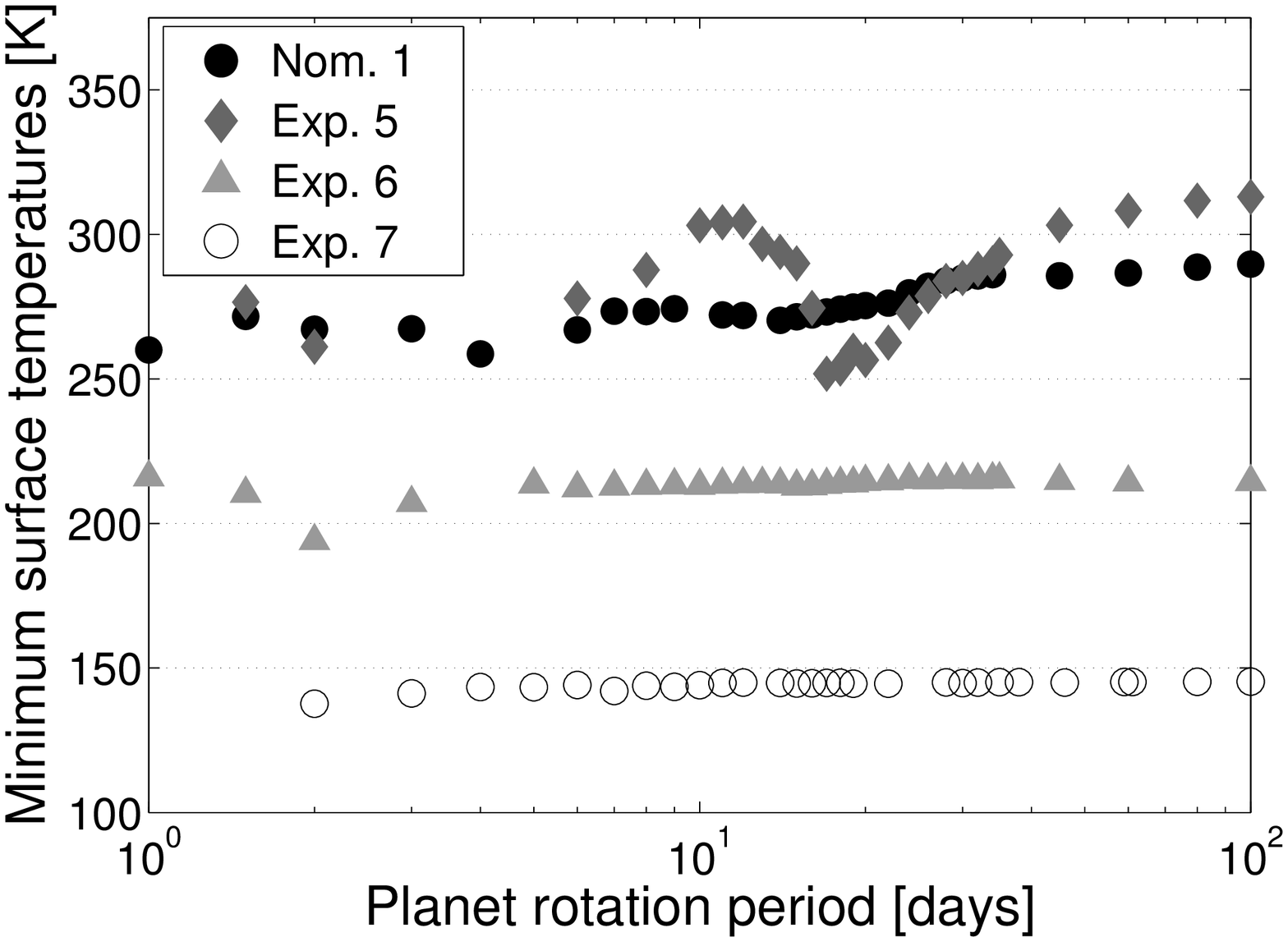}
\caption{Maximum and minimum surface temperature for different surface friction time scales $t_{s, fric}=0.1-100$~days and $R_P=1.45 R_{Earth}$.}
\label{fig: Temp_fric}
\end{figure}

\begin{figure}
\includegraphics[width=0.48\textwidth]{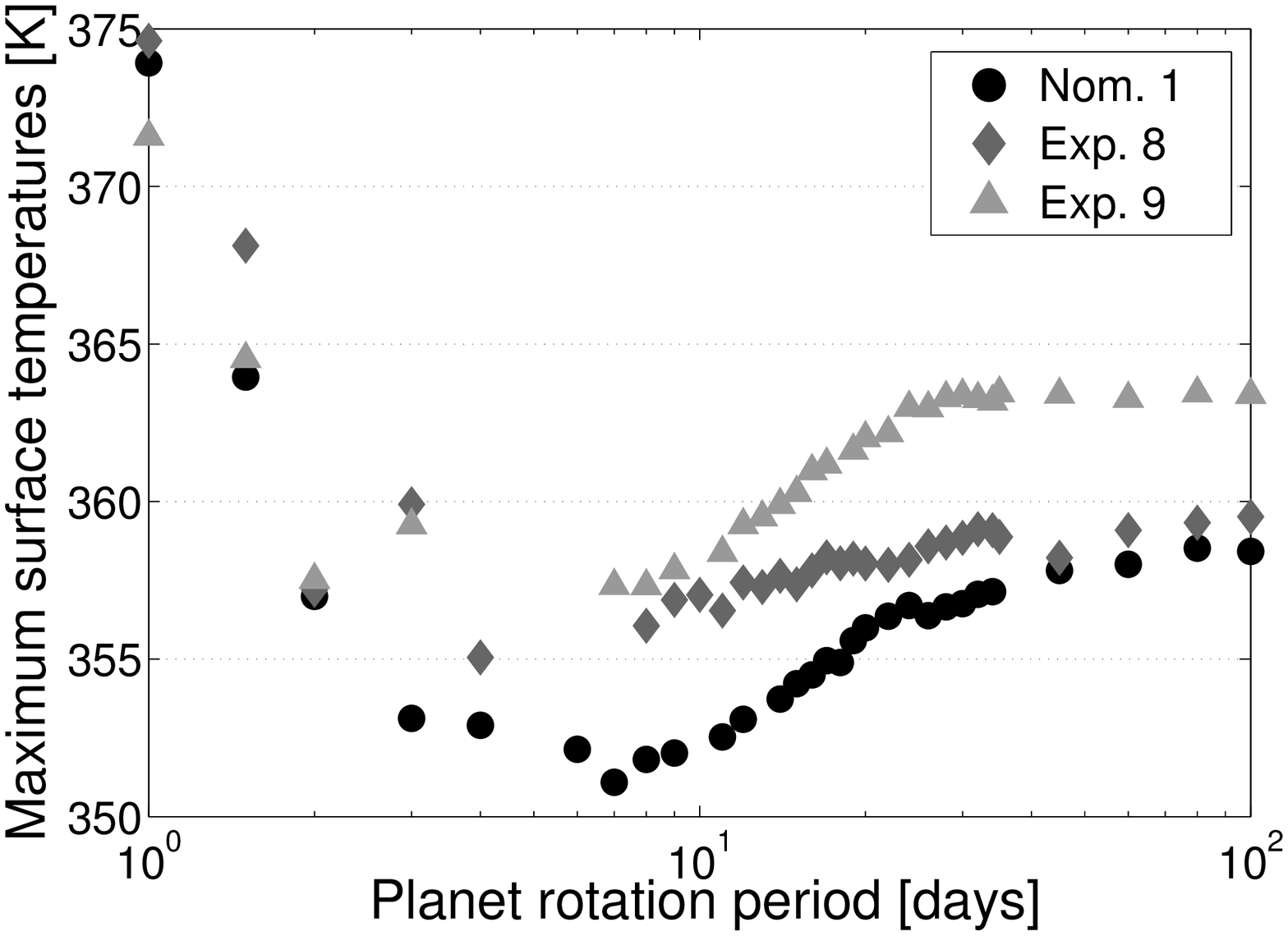}
\includegraphics[width=0.48\textwidth]{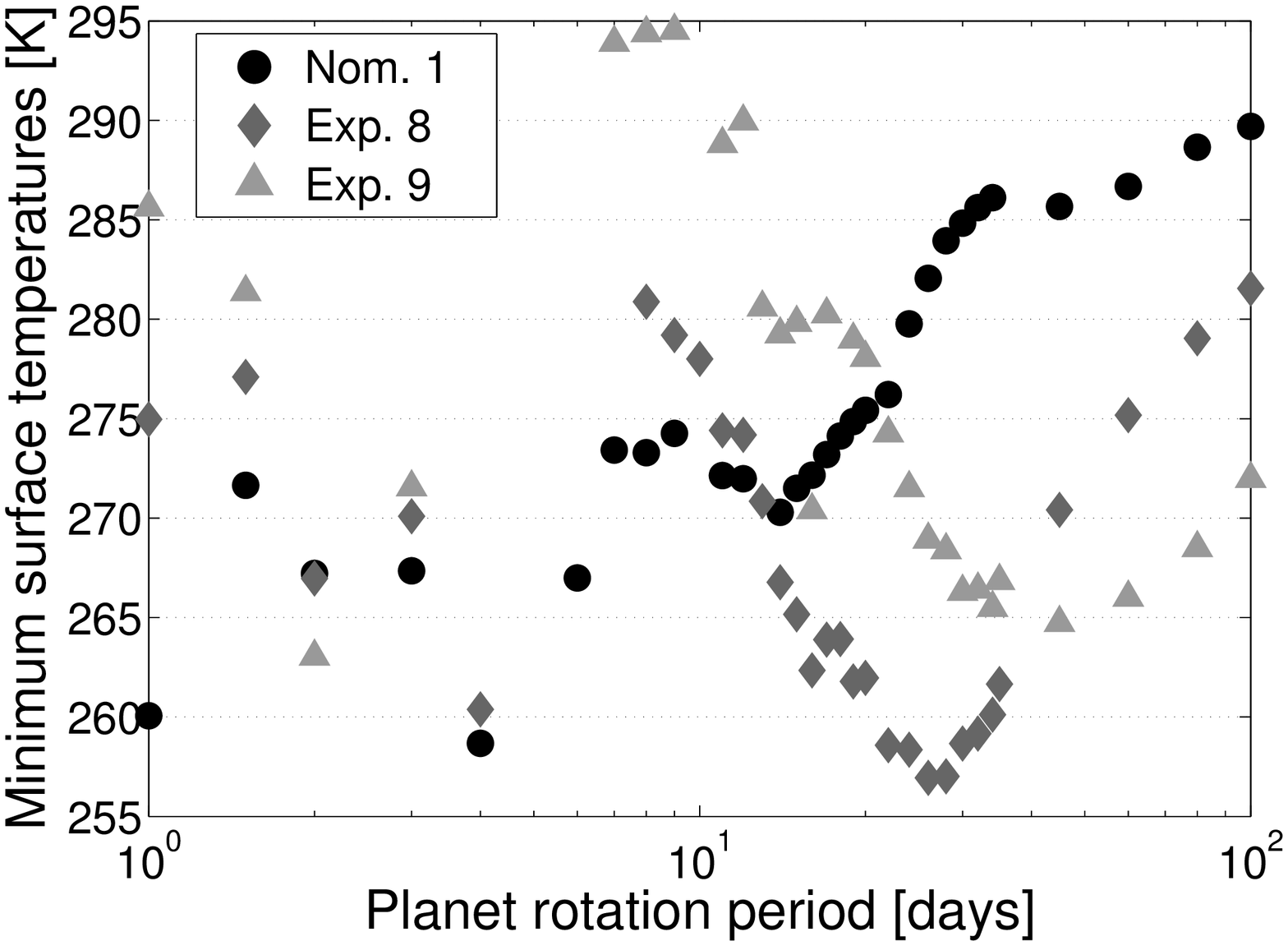}
\caption{Maximum and minimum surface temperature for different PBL extent $p_{PBL}=700-900$~mbar and $R_P=1.45 R_{Earth}$.}
\label{fig: Temp_PBL}
\end{figure}

The surface temperatures in the models with the least efficient surface friction (\emph{Exp.6} and \emph{Exp.7} with $t_{s,fric}$ = 10 and 100 days, respectively) are basically decoupled from atmosphere dynamics. They do not evolve with changes in climate states and circulation structure in the whole rotation period regime (Figure~\ref{fig: Temp_fric}). On the other hand, the surface temperatures exhibit huge differences between the day side and the night side with $\Delta T=$ 110~K and 230~K for \emph{Exp.6} and \emph{Exp.7}, respectively. In particular, the night side is adversely affected by the reduction in heat transfer near the surface and assumes very cold surface temperatures of $T_{NS,s}=$~220 K and 150 K, respectively. At no point, however, do the models reach temperatures that would lead to out-freezing of the primary atmosphere component. The day side surface temperatures are comparatively less affected by the inefficiency of heat transport and reach at the substellar point temperatures of $T_{DS,s}\approx $~380~K. These substellar point temperatures can also be reached for fast rotating tidally locked planets ($P_{rot}=1$~day) if strong equatorial superrotation suppresses cooling of the substellar point via upwelling (\textit{Nom.1}). The very cold night sides were already inferred in Section~4.2.2 from the very deep negative geopotential height perturbation. We thus confirm here that \emph{Exp.6} and \emph{Exp.7} have indeed a particularly large day side to night side temperature gradient $\Delta T$.

Although the surface temperatures are apparently unaffected by different climate states for experiments with weak surface
friction, the very cold night side temperatures and the large temperature gradient influence in turn climate states - in particular for slow and intermediate rotations ($P_{rot}=12-100$~days). The resulting strong negative geopotential anomalies $z'$ lead to stronger Kelvin and Rossby waves already for $P_{rot}=100$~days. For $z'\leq 200$~m or $T=150~$K at the night side in \emph{Exp.7}, we even find a change in zonal wind direction from an equatorial westerly (superrotation) to equatorial easterly flow (anti-rotation).

The surface temperatures of experiments with efficient surface friction $t_{s,fric} \leq$ 1~days (\textit{Exp.5} and \textit{Nom.1}) evolve with planet rotation. Generally, the night side and surface temperatures cool down with faster rotation between $P_{rot}=100$ and 10~days. \textit{Exp.5}, however, deviates between $P_{rot}=10-15$~days from this nominal evolution outlined in C15. As shown previously (Section~\ref{sec: fric_lambda1}), \textit{Exp.5} experiences in this rotation regime a climate state with particularly strong coupling between standing tropical Rossby and Kelvin waves. This strong coupling leads furthermore to a weakening of the direct circulation (see previous Section, Figure~\ref{fig: circ_fric}). At the same time, secondary circulation cells are weak in the relevant rotation regime (Table~\ref{tab: circulation2}). Thus, we would expect warmer substellar point temperatures and colder night side temperatures compared to slower planetary rotation. Surprisingly, the exact opposite is observed (Figure~\ref{fig: Temp_fric}). Either the relative strength of the embedded reverse circulation (\emph{state 1}) provides an alternative heat transport mechanism between $P_{rot}=10-15$~days in \textit{Exp.5} or the overall weakness in the vertically fragmented direct equatorial cells is compensated by more efficient heat transport close to the surface in the lower branch of the direct circulation cell.

The evolution of the day side surface temperatures in the experiments with compressed PBL (\emph{Exp.8} and \emph{Exp.9}) shows generally the same evolution with rotation period as the nominal model (\textit{Nom.1}). The surface temperatures are just warmer by a few Kelvin if the PBL extent is compressed (Figure~\ref{fig: Temp_PBL}, upper panel).

The evolution of the night side surface temperatures, however, is more strongly affected by changes in climate states: The night side surface temperatures show variations of up to 32~K for changes in the PBL extent (Figure~\ref{fig: Temp_PBL}, lower panel). For slow rotators ($P_{rot}=40-100$~days), less heat reaches the night side via direct circulation the more compressed the PBL is.

In the intermediate to slow rotation regime ($P_{rot}=10-40$~days), \textit{Exp.8} and \textit{Exp.9} show a very different night side temperature evolution with faster planet rotation compared to the nominal model (\textit{Nom.1}): The night side surface temperatures rise steeply with faster planet rotation. A new heat transport mechanism develops in this rotation regime that becomes more and more efficient with faster rotation.

One obvious candidate for such a heat transport mechanism is secondary circulation cells (circulation \textit{state 2}) that, indeed, start to develop for surprisingly slow rotations for experiments with compressed PBL and gain in strength with faster rotation (Figure~\ref{fig: circ_PBL}, lower panel and Table~\ref{tab: circulation2}).  Closer comparison between the increase in night side surface temperatures with faster rotation and the development of circulation \textit{state 2} shows, however, that both don't match in rotation period phase space. Circulation \textit{state 2 }forms for $P_{rot} \leq 17-22$~days and the night side surface temperatures increase for $P_{rot}\leq 30$ and 40 days in \emph{Exp. 8} and \emph{9}, respectively (Compare Table~\ref{tab: circulation2}, Figure~\ref{fig: circ_PBL}, lower panel and Figure~\ref{fig: Temp_PBL}, lower panel).

Since direct circulation strength decreases with faster rotation in \emph{Exp.8} and \emph{Exp.9}, we conclude that there is only one mechanism that can explain the warming of the night side with faster rotation in the intermediate to slow rotation regime ($P_{rot}=10-40$~days): the embedded reverse circulation cells of circulation \textit{state 1}. They are already present for $P_{rot}= 30$ and 40 days, respectively (see previous Section) and it was shown in C15 that they also gain in strength with faster planet rotation.

We speculate that the night surface is not warmed in nominal model (\textit{Nom.1}) by embedded reverse cells, because the cells are located above the PBL. $p_{PBL}=700$~mbar is relatively high in \textit{Nom.1} (see C14). If the upper extent of the PBL is lowered, so is the location of the embedded reverse cell. The closer the embedded reverse cell comes to the surface, the more heat it can transport towards the surface's night side.

For fast planet rotation ($P_{rot}\leq 10$~days), the embedded circulation cells don't transport heat any more and the night side surface temperatures of \textit{Exp. 8} and \emph{9} follow again qualitatively the evolution of \textit{Nom.1} with faster planet rotation.

\section{Summary}

We investigated how variations in night side cooling efficiency, surface friction time scales and planetary boundary layer (PBL) height affect climate states of tidally locked Earth-like planets. This is the first study that covers for every scenario the rotation period between $P_{rot}=1-100$~day to also take into account changes in climate states as identified in C15.

We performed a series of nine parametric scans, where we took advantage of the versatility of our model.  First, we varied the thermal forcing:
\begin{itemize}
\item In \emph{Exp.1}, the thermal forcing on the night side of a Super-Earth planet ($R_P=1.45 R_{Earth}$) was increased by gradually reducing the radiative time scale on the night side from $t_{rad,NS}=813$ to $t_{rad,NS}=100$~days.
\item In \textit{Exp.2}, the optical depth was increased from $\tau_s=0.62$ to $\tau_s=0.94$.
\item In \textit{Exp.3}, the planet size was reduced from $R_P=1.45 R_{Earth}$ to $R_P=1 R_{Earth}$, using the same thermal forcing than for \emph{Exp.2}.
\item In \textit{Exp.4}, the radiative time scales on day and night side were decreased by a factor of 1.45 to compensate for increased efficiency of dynamics due to the decrease in planet size between \textit{Exp.2} and \textit{Exp.3}.
\end{itemize}

We find that an increase in night side cooling efficiency lowers the night side surface temperatures significantly by about 80-100~K in \textit{Exp.1}, while the day side surface temperatures are lowered at the same time by 25~K at most.

In our model, an increase in optical depth - keeping all other parameters equal - affects mainly the day side as shown by \textit{Exp.2}: The day side surface temperatures $T_{DS,s}$ rise and the radiative time scale decreases with increasing optical depth, which leads combined to an overall hotter day side despite more vigorous circulation.  In contrast to that a similar experiment conducted by \cite{Joshi1997} showed that an increase in optical depth in their model led to a rise in the night side surface temperatures. \cite{Joshi1997} explicitly state that the radiative time scales in their simplified thermal forcing prescription were unaffected by changes in optical depth. The effect of changes in optical depth can thus only be due to an increase in circulation.

\textit{Exp.3} demonstrated that already a reduction in planet size - keeping all other parameters equal - leads to a reduction of the temperature gradient between the day side and the night side. In particular, the night side is substantially warmer in \textit{Exp.3} compared to \textit{Exp.2}. \textit{Exp.3} thus confirms the results reported by \cite{Joshi1997}: Stronger circulation - keeping radiative time scales equal - leads to a rise in night side surface temperatures.

\textit{Exp.4} demonstrated that the dynamical time scale is scaled to first order with $\sim R_P$. Stronger circulation in \textit{Exp.3} compared to \textit{Exp.2} can thus successfully be compensated by lowering the radiative time scales by a factor of 1.45 to account for reduction in planet size from $R_P=1.45 R_{Earth}$ to $R_P=1 R_{Earth}$. As a result, the day side and night surface temperatures in \textit{Exp.2} and \textit{Exp.4} are similar to each other.

\textit{Exp.4} was found to assume for all intermediate to fast rotation periods $P_{rot}=1 -26$~days a climate state dominated by standing tropical Rossby waves and equatorial superrotation. The nominal model for an Earth-sized planet and with inefficient night side cooling (\textit{Nom.2}) was reported in C15 to also favour climate states with standing tropical Rossby waves in the intermediate rotation regime $P_{rot}=4-26$~days. In the fast planet rotation regime, however, \textit{Nom.1} showed in contrast to \textit{Exp.4} the formation of standing extra tropical Rossby waves with two high latitude westerly jets instead or in addition to an equatorial jet. The latter climate states are called 'mixed states'.

\textit{Exp.4} has stronger thermal forcing than the nominal model for the same planet size with $R_P=1 R_{Earth}$ (\textit{Nom.2}) due to larger optical depth and more efficient night side cooling. We thus conclude that the stronger thermal forcing favours the formation of standing tropical Rossby waves. It suppresses the formation of standing extra tropical Rossby waves for $P_{rot}=1.5-3$~days.  Thus, we have identified strong thermal forcing as one mechanism that drives the model to favour one possible climate state in the short rotation period regime over the other. Despite strong thermal forcing, the model used by \cite{Edson2011} does not suppress standing extra tropical Rossby waves in the fast rotation regime. Thermal forcing can thus only be one possible component that determines which climate state a rocky tidally locked planet assumes in the short rotation period regime.

In the second part of our parameter study, we varied surface boundary treatments. We changed
\begin{itemize}
\item surface friction time scale to $t_{s, fric}=0.1,10$ and 100~days (\emph{Exp.5}, \textit{Exp.6} and \emph{Exp.7}, respectively)
\item and the upper extent of the planet boundary layer to $p_{PBL}=800$ and 900~mbar (\emph{Exp.8} and \emph{Exp.9}, respectively).
\item $R_P=1.45 R_{Earth}$ and the nominal thermal forcing was assumed for surface boundary treatment variations.
\end{itemize}

We found large climate state differences in the whole rotation period regime ($P_{rot}=1-100$~days) for surface friction and boundary layer variations.

Most notably, \emph{Exp. 5} assumes climate states dominated by the extra tropical Rossby wave with two high latitude westerly jets in the fast rotation regime ($P_{rot}\leq 10$~days) - in contrast to all other experiments. For very fast rotations ($P_{rot}\leq $~2 days), \emph{Exp. 5} exhibits mixed climate states with both tropical and extra tropical standing Rossby waves. Therefore, we have identified the second mechanism that determines which climate state a rocky tidally locked planet assumes in the short rotation period regime in addition to strength of thermal forcing: Surface friction efficiency. Apparently, strong surface friction favours the formation of standing extra tropical Rossby waves as soon as the Rossby wave number $L_R/R_P$ becomes smaller than unity. Already \cite{Showman2011} speculated that surface friction may be one mechanism on rocky planets that may give rise to climate dynamics that deviate from tropical Rossby waves exhibited by gas planets.

Interestingly, \emph{Exp. 5} assumes for intermediate rotations ($P_{rot}=10-15$~days), where extra tropical Rossby waves cannot form because $L_R/R_P>1$, a climate state with particularly strong coupling between standing tropical Rossby and Kelvin waves. This coupling leads not only to very strong equatorial winds but also to suppressed direct circulation and strong embedded reverse circulation.

Experiments with weaker surface friction efficiency than the nominal model, \emph{Exp.6} and \emph{7}, show decoupling between the surface temperatures and circulation: The surface temperatures hardly change in the investigated rotation period regime ($P_{rot}=1-100$~days) and the night side surface assumes very cold temperatures ($T_{NS,s}=220$~K and 150~K for \emph{Exp.6} and \emph{7}, respectively). While surface temperatures do not evolve with rotation period and are thus not affected by changes in climate states, the cold night sides, however, affect the possible climate states instead - in particular in the slow and intermediate rotation regime ($P_{rot}=12-100$~days).

Simulations with very cold night side surface temperatures ($T_{NS,s}\leq 220$~K), that is \emph{Exp.6} and also \emph{Exp.4}, trigger in the intermediate to slow rotation regime ($P_{rot}=10-100$~days) stronger tropical Rossby waves, which lead to stronger equatorial superrotation. \emph{Exp.7} with the coldest night side ($T_{NS,s}\approx 150$~K) shows, surprisingly, a drastically different climate state for intermediate to slow planet rotation: \textit{Exp.7} exhibits strong easterly winds ($u=-50$ to $-80$~m/s). They even appear to suppress the formation of tropical Rossby waves with faster planet rotation. The strength of the easterly wind in \emph{Exp.7} diminishes with faster rotation and eventually the wind system displays 'normal' westerly equatorial superrotation - albeit for comparatively fast rotations ($P_{rot} \leq $~12~days). The onset of westerly superrotation is accompanied by the formation of standing tropical Rossby waves. We are currently at a loss to explain the peculiar behaviour of \emph{Exp.7} and tentatively link it to the extremely cold night side exhibited in that experiment.

When the PBL is more compressed (\emph{Exp.8} and \emph{Exp.9}), the simulations develop colder night side surface temperatures than the nominal model in the slow rotation regime. They do not reach, however, the temperatures  of models with very weak surface friction \textit{(Exp.6,7}): They assume $T_{NS,s}=255$ and 265~K, respectively.

\emph{Exp.8} and \emph{Exp.9} - and also \emph{Exp.4} with efficient night side cooling - show, surprisingly, already tropical Rossby wave gyres in the slow rotation regime ($P_{rot}=$ 100~days). We concluded that for very strong thermal forcing in our model prescription, Rossby wave gyres can form already for $\lambda_R/R_P$ larger than unity: $\lambda_R/R_P=1.75-2$. Closer inspection of flow and geopotential perturbations for $\lambda_R/R_P\approx 1$ confirmed, on the other hand, that full formation of tropical Rossby waves takes indeed place solely in the $\lambda_R/R_P < 1$ regime, that is, for $P_{rot} < 34$~days. As a side note, although the dry model used by \cite{Edson2011} appears to have similar strength in thermal forcing as \textit{Exp.4}, they do not report weak Rossby waves for $P_{rot}=100$~days. However, we have already concluded that thermal forcing alone does not determine which planetary wave can form.  At least surface friction has to be taken into account as well. Unfortunately, we don't know how efficient surface friction is in \cite{Edson2011}.

Another surprise was found for the experiments with compressed PBL (\emph{Exp. 8} and \emph{9}) in the intermediate to slow rotation regime $P_{rot}=10-100$~days: The tropical Rossby wave is capable to trigger the formation of embedded reverse cells already for slow rotations. Furthermore, these cells appear to transport efficiently heat from the day side towards the night side between $P_{rot}=10-40$~days. The fact that we only find embedded reverse cells in climate states with tropical Rossby waves and that these are completely suppressed otherwise, e.g. in \textit{Exp.7} for $P_{rot}=12-100$~days, confirms the conclusion in C15 that these features are exclusively linked to tropical planetary waves.

\section{Conclusion and outlook}

This study provides a coherent link between climate states and basic assumptions in thermal forcing and frictional solid surface boundaries. Due to the computational efficiency of our model, we could perform many experiments (ca. 700 Experiments including C15) that cover the relevant rotation period $P_{rot}=1-100$~days in fine detail. Our study provides a better understanding of how surface temperatures and circulation can vary between different climate models for tidally locked terrestrial planets. This understanding is particularly important for rocky planets on relatively tight orbits, whose atmosphere properties will become first accessible to observations.

We found that strong surface friction ($t_{s, fric}=0.1$~days) promotes the formation of extra tropical Rossby waves for fast rotating terrestrial planets ($P_{rot} \leq$ 10 days), as already noted by \cite{Showman2011}. This rotation period region is of relevance for the extended inner habitable zone that ranges at least down to $P_{rot}=6$~days for M dwarf stars (\cite{Zsom2013,Seager2013}). It is also of relevance for the model of \cite{Yang2014} who found that the inner edge of the habitable zone can be expanded towards the star for tidally locked planets due to cloud coverage over the substellar point. Cloud formation over the substellar point is, however, only possible with unperturbed upwelling over the substellar point and thus unperturbed direct circulation cell. The latter requires climate states that are dominated by standing extra tropical Rossby waves in the fast rotation regime. Otherwise, equatorial superrotation brought about by the formation of standing tropical Rossby waves would disrupt direct circulation. Thus, as discussed in C15, different circulation states in the short rotation regime yield different prospects on habitability.

Surface friction alone, however, is not the only mechanism that determines which planetary wave and thus climate state can form on a tidally locked rocky planet with fast rotation. Also thermal forcing was found to play a strong role: Stronger thermal forcing, e.g. due to very efficient night side cooling, was found to lead to a climate state dominated rather by tropical standing Rossby waves instead of extra tropical Rossby waves.

In addition, not only surface friction in terrestrial planets but also friction in gas planets can, apparently, excite climate states that are not exclusively dominated by equatorial superrotation: \cite{Kataria2015} report for the Hot Jupiter WASP-43b an evolution from a climate state with an equatorial superrotating jet to a mixed state with two westerly high latitude jets and an equatorial jet, if frictional drag is uniformly increased from $t_{fric}=10^6$~s to $10^5$~s. Thus, our study can be linked to the study of Hot Jupiters.

We will, therefore, develop our model further to also encompass the highly interesting hot Super-Earths in the intermediate region between terrestrial and gas planets. We will investigate if planets with a solid surface are indeed in a different climate state than planets without a solid boundary and if climate state differences could be potentially observed, e.g. by the upcoming James Webb Space Telescope (JWST).

We will address furthermore how ambiguities can be solved: E.g. how to disentangle surface drag from other drag mechanisms that may be present in gas planets. Furthermore, \cite{Kataria2014} showed surprisingly that even changes in atmosphere composition can induce a change in climate state. Atmospheres with high molecular mean weight like CO${}_2$ can like-wise exhibit a mixed state with high latitude jets and a weak equatorial jet. The authors compared, however only few simulations. We plan, therefore, to explore the hot Super-Earth climate regime with more experiments to resolve ambiguities and to investigate the reliability of model results to pave the way for the observations of smaller exoplanets.

\section*{Acknowledgments}
We acknowledge support from the KU Leuven projects IDO/10/2013 and GOA/2015-014 (2014-2018 KU Leuven). The computational resources and services used in this work were provided by the VSC (Flemish Supercomputer Center), funded by the Hercules Foundation and the Flemish Government – department EWI. Furthermore, we wish to thank the anonymous referee whose insightful remarks and suggestions led to a substantial improvement in the quality of this paper.

\bibliography{GCM} 
\label{lastpage}
\end{document}